\RequirePackage{fix-cm}
\documentclass[smallextended]{svjour3}       
\smartqed  
\usepackage{graphicx}
\usepackage{graphics}
\usepackage{lineno}
\usepackage{hyperref}
\modulolinenumbers[5]
\usepackage{amsmath,amssymb,fixmath}
\usepackage{epsfig,epstopdf}
\usepackage{booktabs}
\usepackage{multirow}
\usepackage{subfig}
\usepackage{float}
\usepackage{mwe}
\usepackage{graphicx}
\usepackage{caption}
\usepackage{url}

\usepackage{array}
\usepackage{soul}
\usepackage{xcolor}

\usepackage[misc]{ifsym}

\usepackage{enumitem}
\usepackage{booktabs}
\usepackage{longtable}
\usepackage{stmaryrd}
\usepackage{subfloat}
\usepackage{tabularx}
\usepackage{pdflscape}
\newcommand{\ds}{\displaystyle}
\usepackage[square,sort,comma,numbers]{natbib}

\begin{document}

\title{A Novel Quantum Calculus-based Complex Least Mean Square Algorithm ($q$-CLMS)}

\titlerunning{$q$-CLMS}        

\author{Alishba~Sadiq \and 
Imran~Naseem \and
Shujaat~Khan \and
Muhammad~Moinuddin \and
Roberto~Togneri \and
Mohammed~Bennamoun}


	\institute{
	    A.Sadiq \at
		Centre for Intelligent Signal and Imaging Research (CISIR), Department of Electrical and Electronic Engineering, Universiti Teknologi PETRONAS, Malaysia
		 \\
		\email{alishba\_18001778@utp.edu.my}  
		\and
		I.Naseem \Letter \at
		College of Engineering, Karachi Institute of Economics and Technology, Korangi Creek, Karachi 75190, Pakistan.  \\School of Electrical, Electronic and Computer Engineering, The University of Western Australia, 35 Stirling Highway, Crawley, Western Australia 6009, Australia.\\
		\email{imran.naseem@uwa.edu.au}
		\and
		R.Togneri \at
	    Department of Electrical, Electronic and Computer Engineering, The University of Western Australia, 35 Stirling Highway, Crawley, Western Australia 6009, Australia. \\
		\email{roberto.togneri@uwa.edu.au}
		\and 
		S.Khan \at 
		Department of Bio and Brain Engineering, Korea Advanced Institute of Science and Technology (KAIST), Daejeon, Republic of Korea.\\
		\email{shujaat@kaist.ac.kr}
		 \and
		 M.Moinuddin\at
        Center of Excellence in Intelligent Engineering Systems (CEIES),  King Abdulaziz University, Saudi Arabia.\\
        \email{mmsansari@kau.edu.sa}
        \and
        M.Bennamoun \at
        Department of Computer Science and Software Engineering, The University of Western Australia, 35 Stirling Highway, Crawley, Western Australia 6009, Australia.\\
        \email{mohammed.bennamoun@uwa.edu.au}}
        
        \date{Received: date / Accepted: date}

\date{Received: date / Accepted: date}

\maketitle

\begin{abstract}
In this research, a novel adaptive filtering algorithm is proposed for complex domain signal processing. The proposed algorithm is based on Wirtinger calculus and is called as \emph{q}-Complex Least Mean Square (\emph{q}-CLMS) algorithm.  The proposed algorithm could be considered as an extension of the $q$-LMS algorithm for the complex domain.  Transient and steady-state analyses of the proposed $q$-CLMS algorithm are performed and exact analytical expressions for mean analysis, mean square error (MSE), excess mean square error (EMSE), mean square deviation (MSD) and misadjustment are presented.  Extensive experiments have been conducted and a good match between the simulation results and theoretical findings is reported.  The proposed $q$-CLMS algorithm is also explored for whitening applications with satisfactory performance. A modification of the proposed $q$-CLMS algorithm called Enhanced $q$-CLMS (E$q$-CLMS) is also proposed.  The E$q$-CLMS algorithm eliminates the need for a pre-coded value of the $q$-parameter thereby automatically adapting to the best value.  Extensive experiments are performed on system identification and channel equalization tasks and the proposed algorithm is shown to outperform several benchmark and state-of-the-art approaches namely Complex Least Mean Square (CLMS), Normalized Complex Least Mean Square (NCLMS), Variable Step Size Complex Least Mean Square (VSS-CLMS), Complex FLMS (CFLMS) and Fractional-ordered-CLMS (FoCLMS) algorithms.
\keywords{ Complex Least Mean Squares Algorithm, steady-state analysis, transient analysis, $q$-calculus, $q$-LMS, Wirtinger derivative.}
\end{abstract}
\section{Introduction}\label{Sec:Intro}
Adaptive algorithms have been successfully applied in a number of signal processing applications such as channel equalization, noise cancellation, system identification and optimization, etc \cite{NCLMS,AI3,VanKien2021}.  Adaptive learning frameworks have also been traditionally used for regression, classification \cite{Ibrokhimov2021} and dimensionality reduction \cite{Wu2021,Chen2021} encompassing several applications such as image segmentation \cite{Ding2021}, quality enhancement \cite{Li2021}, etc.

The Least Mean Square (LMS) method is considered to be a cornerstone approach with several merits and is one of the most widely used adaptive algorithms. Because the LMS is dependent on the eigenvalue spread of the input correlation matrix, it suffers from slow convergence. Several adaptive strategies, such as normalized LMS (NLMS), computed adaptive learning rates, a chaotic teaching-learning based optimization, variable power fractional LMS algorithm, and variable step-size LMS algorithm, have been presented in the literature to address this issue, \cite{NLMS,AI1,VPFLMS,RVPFLMS,RVSSFLMS,VSS}.

Recently, an interesting variant of the traditional LMS algorithm, called the q-LMS \cite{qLMS}, has been proposed.  The q-LMS algorithm computes the derivative of the cost function using secant rather than tangent function as it utilizes the notion of $q$-gradient \cite{qLMS}.  The method is shown to achieve a higher convergence rate, compared to the conventional LMS, while maintaining competitive performance.  In \cite{8955461,qNLMS,mishra2021q,hussain2021q,EqLMS,sadiq2020q}, adaptive techniques are further proposed for the $q$ parameter. The \emph{q}-LMS algorithm also been successfully implemented for system identification, unconstrained optimization, neural networks and the design of whitening filters tasks.

Typically, adaptive filters are used with real-valued signals and systems.  For signals with intensity and direction components, complex domain provides an efficient framework \cite{direction}.  As adaptive algorithms requires the minimization of a cost function through differentiation, wirtinger calculus is used for the evaluation of a cost function in complex domain.  In a number of  fields such as beamforming, object detection, bio-medicine and communications, complex modeling of signals and systems is required  \cite{khan2020adaptive,radar,biomed}.  A number of applications of Wirtinger calculus can be found in areas such as high performance distribution networks, complex-valued neural networks and complex kernel LMS\cite{powerflow,CVNN,CKLMS}.  The analysis of such systems is not straightforward.  Simplified approaches, such as the processing of the imaginary and real parts separately, tend to limit the usefulness of the complex domain.  The cornerstone work in this regard is the CLMS (Complex Least Mean Square) \cite{CLMS} with the weight update equation:
\begin{equation}\label{CLMS}
\mathbf{w}(n+1) = \mathbf{w}(n) +2\mu \mathbf{x}(n)e^{*}(n).
\end{equation}
where $\mu$ is the convergence factor, $e^{*}(n)$ is the conjugate of the error, and $\mathbf{w}(n)$ and $\mathbf{w}(n+1) $ are the current and updated weights, respectively.  The CLMS algorithm has been applied in various fields, such as adaptive prediction, system identification and complex dual-channel estimation. There are some existing works in the literature to improve the CLMS algorithm and various variants have been proposed \cite{LMS9,LMS11,FCLMS,RNACLMS,Shah2017,Ahmad2021}.  In this study, we propose a quantum variant of the complex least mean square  by utilizing the novel \emph{q}-wirtinger calculus based \emph{q}-steepest decent method. To the best of our knowledge, this is for the first time where a quantum calculus-based approach has been applied to design an adaptive learning method for the complex domain signal processing applications.  The proposed algorithm is called as the $q$-CLMS ($q$-Complex Least Mean Square) algorithm.
Major contributions of the proposed research are as follows:
\begin{itemize}
	\item[$\bullet$]  A novel  \emph{q}-calculus-based adaptive filtering algorithm is proposed for the complex domain (section \ref{Sec:Pro}). In particular, the notion of the $q$-LMS is extended to the complex domain using the Wirtinger calculus. \cite{ASP}.
	\item[$\bullet$] Extensive theoretical analysis of the proposed \emph{q}-CLMS algorithm, consisting of the mean and the mean-square error convergence, is performed. Learning curves for Excess Mean Square Error (EMSE), Mean Square Deviation (MSD), and misadjustment is also derived.
	\item[$\bullet$] The proposed $q$-CLMS is further improved by making the $q$-parameter time-varying (section \ref{Sec:time-varying}).  This modified approach eliminates the need for a pre-coded value of the $q$-parameter by automatically adapting to the best value.  The modified approach is called Enhanced \emph{q}-CLMS (E\emph{q}-CLMS).
	\item[$\bullet$] For the system identification task, compact analytical forms have been derived and validated through the simulation results (sub-section \ref{valid}).
	\item[$\bullet$] For the selection of $q$-parameter, the \emph{q}-gradient is applied in such a manner as to make the LMS filter perform as a whitening filter.
	\item[$\bullet$] Extensive comparative analysis has been carried out for the proposed algorithm where results are compared to several state-of-the-art approaches namely the complex LMS (CLMS), normalized CLMS (NCLMS) and variable step-size CLMS (VSS-CLMS).  The experiments clearly show the efficacy of the proposed approach.  
	\item[$\bullet$] To show the efficacy of the proposed method, channel equalization task has been addressed and the comparison has been made with the fractional variants of the LMS namely, CFLMS \cite{Shah2017} and FoCLMS \cite{Ahmad2021}.
\end{itemize}

The rest of the paper is composed as follows:  In Section \ref{Sec:Overview}, \emph{q}-calculus and the $q$-steepest descent algorithm are explained.  The proposed \emph{q}-CLMS algorithm is discussed with detailed analysis in Section \ref{Sec:Pro}.  The proposed $q$-CLMS algorithm is modified to an adaptive version i.e., E$q$-CLMS in Section \ref{Sec:time-varying}.  An interesting application of the proposed methodology namely the whitening filter is discussed in Section  \ref{Sec:whitening} followed by extensive experiments and results in Section \ref{Sec:Sim}.  The paper is finally concluded in Section \ref{Sec:Con}.

\section{The Proposed \emph{q}-Wirtinger calculus}\label{Sec:Overview}

Quantum calculus is often specified as the calculus with no limits \cite{qbook1}.  It has been applied in numerous fields and a collection of new outcomes can be found in \cite{qbook2,qbook3,qbook4}.

The differential of a function in quantum calculus is given as \cite{kac2001quantum}:
\begin{equation}
d_{q}(f(x)) = f(qx)-f(x).
\end{equation}
Consequently, the derivative of a function is defined as
\begin{equation}\label{q}
D_{q}(f(x)) = \frac{d_{q}(f(x))}{d_{q}(x)} = \frac{f(qx)-f(x)}{(q-1)x}.
\end{equation}
The formula takes the form of the standard derivative as $\emph{q}$ approaches 1.
For the function  $x^{n}$, the \emph{q}-derivative is given as:
\begin{equation}\label{q_derivative1}
D_{q,x}x^{n} =\begin{cases}
\ds\frac{q^{n}-1}{q-1} x^{n-1}, & q\neq1,
\\ 
\ds nx^{n-1}, & q = 1.
\end{cases}
\end{equation}
The \emph{q}-gradient of a function $f(\mathbf{x})$, for $n$ number of variables, $\mathbf{x}(n) = [{x}_{1}(n),{x}_{2}(n), \dots {x}_{M}(n)]^{\intercal}$, is defined as \cite{nthpower}
\begin{equation}\label{q_gradient1}
\nabla_{\mathbf{q,x}} f(\mathbf{x})\triangleq [D_{q1,x1}f(\mathbf{x}),D_{q2,x2}f(\mathbf{x}),...D_{qn,xn}f(\mathbf{x})]^{\intercal} ,
\end{equation}
Alternatively, $\nabla_{\mathbf{q,x}} f(\mathbf{x})$ can also be written as:
\begin{equation}
\nabla_{\mathbf{q}}f(\mathbf{x})(\mathbf{x})\triangleq [D_{q1}f(\mathbf{x})(x_{1}),D_{q2}f(\mathbf{x})(x_{2}),...D_{q}f(\mathbf{x})(x_{n})]^{\intercal} ,
\end{equation}
where $\mathbf{q}(n) = [{q}_{1}(n),{q}_{2}(n), \dots {q}_{M}(n)]^{\intercal}$.\\
\emph{q}-calculus is a well-established field with well-defined rules \cite{qbook1}. 
For ordinary derivatives, the chain rule is as follows:\\ Let $g(x)=cx^{k}$.  Then
\begin{eqnarray}\label{chainrule}
D_{q} f(g(x)) = D_{q}^{k}(f)(g(x))D_{q}(g)(x).
\end{eqnarray}
The gradient descent methodology is employed in a number of adaptive algorithms, and the $q$-calculus has been employed to develop the $q$-gradient descent for real systems by calculating the secant of the cost-function. (Section \ref{Sec:Intro})
\subsection{\emph{q}-Steepest Descent}
The Weight-update equation of the conventional steepest descent algorithm is given as:
\begin{eqnarray}\label{gradient}
\mathbf{w}(n+1) =\mathbf{w}(n) - \mu \nabla_{\mathbf{w}}J(\mathbf{w}).
\end{eqnarray}
where $\mu$ denotes the step size, $\mathbf{w}(n)$ and $\mathbf{w}(n+1)$ denote the current and updated weights, and $J(\mathbf{w})$ denotes the steepest descent algorithm's cost function, which is given as:
\begin{equation}\label{nonanalytic}
J(\mathbf{w}) = e^{2}(n)
\end{equation}
The estimation error, denoted by the symbol $e(n)$, is defined as the difference between the desired and the output signal. It is written as follows:
\begin{equation}
e(n) = d(n) - y(n)
\end{equation}
Replacing the conventional gradient in Eq.\eqref{gradient} by the $q$ derivative:
\begin{eqnarray}
\mathbf{w}(n+1) =\mathbf{w}(n) - \mu \nabla_{\mathbf{q,w}}J(\mathbf{w}).
\end{eqnarray}
Now, by using Eqs.\eqref{q_derivative1} and \eqref{q_gradient1}, the $\nabla_{\mathbf{q,w}}J(\mathbf{w})$ is calculated as \cite{qLMS}
\begin{equation}\label{q_grad}
\nabla_{\mathbf{q,w}}J(\mathbf{w}) = -2E[\mathbf{G}\mathbf{x}(n)e(n)]
\end{equation}
where $\mathbf{G}$ is defined as a diagonal matrix and the $l^{th}$ diagonal entry of this $\mathbf{G}$ matrix is \\$g_{l} = \frac {q_{l}+1} {2}$, that is\\
\begin{eqnarray}\label{G}
diag(\mathbf{G}) = [(\frac{q_{1}+1}{2}), (\frac{q_{2}+1}{2}),.....(\frac{q_{M}+1}{2})]^{\intercal}
\end{eqnarray}
Substituting $e(n)$ in Eq.\eqref{q_grad} results in the weight update equation of the q-steepest descent algorithm:
\begin{equation}
\mathbf{w}(n+1) = \mathbf{w}(n)  +2\mu \mathbf{G}[\mathbf{P}-\mathbf{R_{x}}  \mathbf{w}(n) ]
\end{equation}
where $\mathbf{R_{x}}$ is the input signal autocorrelation matrix  and $\mathbf{P}$ is defined as the cross correlation matrix between the desired signal $d(n)$ and the input vector $\mathbf{x}(n)$.\\
Considering the successful implementation of the $q$-gradient descent for real systems, we propose to explore the same concept of analysis in the complex domain. The cost-function of the complex LMS algorithm is non-analytic (Eq.\eqref{nonanalytic}). The real valued signals are not analytic, that is, $f : \mathbb{C} \rightarrow \mathbb{R}$ having the imaginary part $v(x, y) = 0$ \cite{ASP}.  The optimization of a cost function poses a research challenge as the cost function are real valued and it can be done by considering the imaginary and real parts of the function separately.  Thus, for simple evaluation of the derivatives in the complex domain, wirtinger calculus provides a suitable framework.  We therefore propose to use Wirtinger method \cite{ASP} to proceed with the $q$-gradient descent.
\subsection{Wirtinger Calculus}
Adaptive algorithms are primarily based on the minimization of a cost function through differentiation.  Cost functions are real valued in the least square problems, hence they are not differentiable in the complex domain. Conventionally,  evaluation of derivatives for the imaginary and real parts are considered separately and then combined \cite{ASP}.  Another suitable concept to evaluate the cost functions directly without separating the imaginary and real parts is known as the Wirtinger calculus \cite{TUM}.  It makes the derivation and analysis much simpler than the conventional splitting approach.\\
Consider a complex function $f(z)$:
\begin{equation}
f(z) = u(x,y)+ j v(x,y)
\end{equation}
Let $f: \mathbb{R}$ x $\mathbb{R}$ $\Rightarrow$ $\mathbb{C}$ be a function of real variables $x$ and $y$ such that $f(z, z^{*}) $= $f(x, y)$, where $z=x +jy$ and that $f(z,z^{*})$ is analytic with respect to $z^{*}$ and $z$ independently. Then,
to derive the form of the differential operators, we write the two real-variables as:
\begin{eqnarray}
x = \frac{z+z^{*}}{2}& and & y = \frac{z-z^{*}}{2j}
\end{eqnarray}
where $z=x+ j y$ and $z^{*}=x- j y$\\
The chain rule to derive the two Wirtinger partial derivative operators for $f(z)$ gives \cite{ASP}:
\begin{eqnarray}\label{W1}
\frac{\partial f}{\partial z} = \frac{\partial f}{\partial x} \frac{\partial x}{\partial z}+\frac{\partial f}{\partial y}\frac{\partial y}{\partial z}\\ = \frac{\partial f}{\partial x} \frac{1}{2}+\frac{\partial f}{\partial y}\frac{1}{2j}
\end{eqnarray}
and
\begin{eqnarray}\label{W2}
\frac{\partial f}{\partial z^{*}} = \frac{\partial f}{\partial x} \frac{\partial x}{\partial z^{*}}+\frac{\partial f}{\partial y}\frac{\partial y}{\partial z^{*}}\\ = \frac{\partial f}{\partial x} \frac{1}{2}-\frac{\partial f}{\partial y}\frac{1}{2j}
\end{eqnarray}

The key point is to treat $z$ and $z^{*}$ independently from each others, allowing us to use Wirtinger calculus.\\

\subsection{The Proposed \emph{q}-Wirtinger Calculus}
We propose to extend the concept of $q$-gradient descent for complex systems using the Wirtinger method in this study.  The proposed methodology results in the inception of a novel calculus called $q$-Wirtinger calculus.\\
Replacing the conventional derivative with the $q$-derivative in Eq.\eqref{W1}:
\begin{equation}
(D_{q}f)(z) = (D_{q}f)(x) (D_{q}x)(z) + ((D_{q}f))(y) (D_{q}y)(z)
\end{equation}
which can be reduced to:
\begin{equation}\label{qW1}
(D_{q}f)(z) = \frac{1}{2}(D_{q}f)(x)  + \frac{1}{2 j }(D_{q}f)(y)
\end{equation}
and
\begin{equation}
(D_{q}f)(z^{*}) = (D_{q}f)(x) (D_{q}x)(z^{*}) + ((D_{q}f))(y) (D_{q}y)(z^{*})
\end{equation}

\begin{equation}\label{qW2}
(D_{q}f)(z^{*}) = \frac{1}{2}(D_{q}f)(x)  - \frac{1}{2 j }(D_{q}f)(y)
\end{equation}
Eqs.\eqref{qW1} and \eqref{qW2} show the $q$-based Wirtinger derivatives.

\section{The Proposed \emph{q}-Complex Least Mean Square Algorithm}\label{Sec:Pro}
The spread of eigenvalues in the input correlation matrix affects the LMS algorithm's performance.  Therefore it is considered as a slow converging method \cite{VSS}.  Taking this problem into consideration, the $q$-LMS has been proposed for real systems in \cite{qLMS}.  Dissimilar to the conventional gradient, the $q$-LMS is derived by making use of the $q$-calculus through the Jackson's derivative approach \cite{ASP}.  This study is based on the $q$-Wirtinger calculus and we propose a novel $q$-complex LMS algorithm.  
Taking the $q$-wirtinger derivative instead of  conventional gradient in Eq.\eqref{gradient} we get:
\begin{eqnarray}\label{q_gradient}
\mathbf{w}(n+1) = \mathbf{w}(n) -\mu (\nabla_{q}J(\mathbf{w}))(\mathbf{w}^{*}(n))
\end{eqnarray}
Using Eq.\eqref{qW2} $D_{q}J(\mathbf{w})(\mathbf{w}^{*}(n))$ can be given as:
\begin{eqnarray}\label{qwirtinger}
\begin{split}
&\nabla_{q}J(\mathbf{w})(\mathbf {w}^{*}(n)) =\\
&\frac{1}{2} \Bigg\{\nabla_{q}J(\mathbf{w})(re[\mathbf{w}^{*}(n)])+ j \nabla_{q}J(\mathbf{w})(im[\mathbf{w}^{*}(n)]) \Bigg\}
\end{split}
\end{eqnarray}
Now by using the chain rule defined in Eq.\eqref{chainrule} the $q$-gradient of the cost function $J(\mathbf{w})$ with respect to $\mathbf{w}^{*}(n)$ is given as
\begin{eqnarray}
\begin{split}
&\nabla_{q}J(\mathbf{w})(re[\mathbf{w}^{*}(n)]) =\\ &\nabla_{q}J(\mathbf{w})(e(n))\nabla_{q}e(n)(y(n))\nabla_{q}y(n)(re[\mathbf{w}^{*}(n)])
\end{split}
\end{eqnarray}
For the first coefficient
\begin{eqnarray}
\nabla_{q1}y(n)(re[w_{1}^{*}(n)]) = -E[(q_{1}+1)e^{*}(n)x_{1}(n)]
\end{eqnarray}
Similarly from 1 to the $M_{th}$ coefficients
\begin{eqnarray}\label{qq_derivative2}
\begin{split}
&\nabla_{q}J(\mathbf{w})(re[\mathbf{w}^{*}(n)]) =\\ &-E[(q_{1}+1)e^{*}(n)x_{1}(n),(q_{2}+1)e^{*}(n)x_{2}(n),\\
&.......(q_{M}+1)e^{*}(n)x_{M}(n)]
\end{split}
\end{eqnarray}
\begin{eqnarray}\label{qq_gradient2}
\nabla_{q}J(\mathbf{w})(re[\mathbf{w}^{*}(n)]) = -2E[\mathbf{G}_{R} \mathbf{x}(n) e^{*}(n)]
\end{eqnarray}

\begin{figure}[!ht]
 	\centering
 	\includegraphics[scale=0.5]{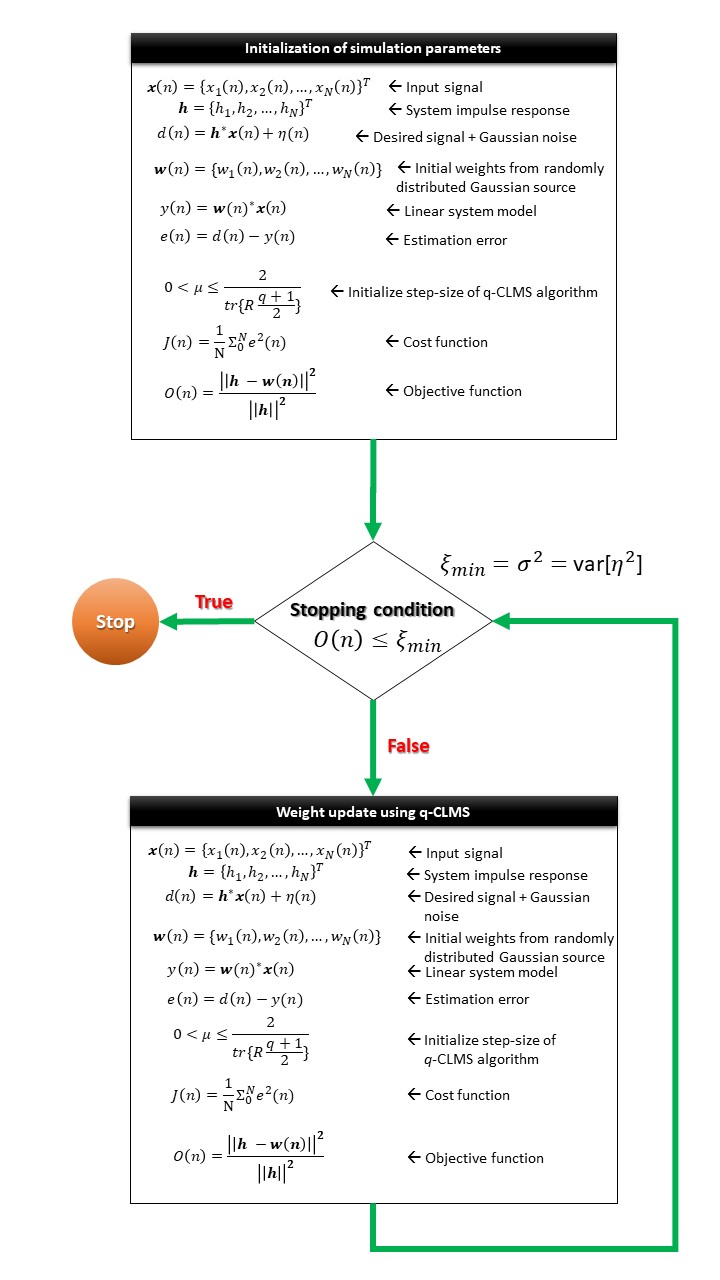}
 	\caption{Proposed q-CLMS algorithm-based system identification flowchart} 
 	\label{flowchart}
 \end{figure}

Similarly for $\nabla_{q}J(\mathbf{w})(im[\mathbf{w}^{*}(n)])$, it is given as:
\begin{eqnarray}
\nabla_{q}J(\mathbf{w})(im[\mathbf{w}^{*}(n)]) = 2E[j \mathbf{G}_{I} \mathbf{x}(n) e^{*}(n)]
\end{eqnarray}
Substituting the value of $\nabla_{q}J(\mathbf{w})(re[\mathbf{w}^{*}(n)])$ and $\nabla_{q}J(\mathbf{w})(im[\mathbf{w}^{*}(n)]) $ in Eq.\eqref{qwirtinger} we get:
\begin{eqnarray}
\begin{split}
&\nabla_{q}J(\mathbf{w})(\mathbf{w}^{*}(n))=\\
&-E[\mathbf{G_{R}}\mathbf{x}(n)e^{*}(n)+\mathbf{G_{I}}\mathbf{x}(n)e^{*}(n)]
\end{split}
\end{eqnarray}
\begin{eqnarray}\label{qwirtinger1}
\nabla_{q}J(\mathbf{w})(\mathbf{w}^{*}(n)) = -E[(\mathbf{G_{R}}+\mathbf{G_{I}}) \mathbf{x}(n) e^{*}(n)]
\end{eqnarray}
Assuming the ergodic nature of the system and hence eliminating the expectation from the \emph{q}-gradient in Eq.\eqref{qwirtinger1} takes the form
\begin{eqnarray}\label{qwirtinger2}
\nabla_{q}J(\mathbf{w})(\mathbf{w}^{*}(n)) \approx -(\mathbf{G_{R}}+\mathbf{G_{I}}) \mathbf{x}(n) e^{*}(n)
\end{eqnarray}
Substituting Eq.\eqref{qwirtinger2} in Eq.\eqref{q_gradient} gives the rule of weight update for the $q$-CLMS algorithm 
\begin{eqnarray}\label{qLMS_final}
\mathbf{w}(n+1) = \mathbf{w}(n) +\mu (\mathbf{G_{R}}+\mathbf{G_{I}}) \mathbf{x}(n)e^{*}(n).
\end{eqnarray}
where $\mathbf{G_{I}}$ and $\mathbf{G_{R}}$ are the imaginary and real parts of the $\mathbf{G}$ matrix, which comprises of \emph{q} dependent entries.  It provides additional control over the conventional CLMS algorithm.\\
The implementation flowchart is shown in Fig.\ref{flowchart}

\subsection{Optimal Solution of the \emph{q}-Complex LMS}
We explore the optimal solution for $q$-CLMS in this sub-section.  As $q$ approaches 1, the
$q$-derivative transforms to the standard derivative. \\
For the optimal solution we replace $\nabla_{q}J(\mathbf{w})(\mathbf{w}^{*}(n))$ in Eq.\eqref{qwirtinger1} with zero.
\begin{equation}\label{opt}
-E[(\mathbf{G_{R}}+\mathbf{G_{I}})\mathbf{x}(n)e^{*}(n)] \approx 0
\end{equation}
By applying the expectation on Eq.\eqref{opt} and replacing $e^{*}(n)$ in Eq.\eqref{opt} we get:
\begin{eqnarray}
(\mathbf{G_{R}}+\mathbf{G_{I}})[\mathbf{P}-\mathbf{R_{x}}\mathbf{w}_{opt}]\approx 0
\end{eqnarray}
where $\mathbf{R_{x}}$ is the auto-correlation matrix of the input signal, and $\mathbf{P}$ is defined as  the cross-correlation vector of the desired output and input.  The optimal weight $\mathbf{w}_{opt}$ is given as:
\begin{equation}
\mathbf{w}_{opt} \approx \mathbf{R_{x}}^{-1}\mathbf{P}
\end{equation}
However, the optimal solution is same as the wiener solution.

\subsection{Convergence of the Mean}\label{Sec:Mean}
We make the following assumptions for the convenience of analysis:
\begin{description}
	\item[$1.$]  We assume that the real part $\mathbf{x}_{R}(n)$ and the imaginary part $\mathbf{x}_{I}(n)$ are uncorrelated, i.e., $E[\mathbf{x}_{R}(n)\mathbf{x}_{I}(n)]$ = $E[\mathbf{x}_{I}(n)\mathbf{x}_{R}(n)]$ = 0.  The input signal auto-correlation matrix is given as:
	$\mathbf{R_{x}} = E[\mathbf{x}_{R}(n)\mathbf{x}_{R}^{\intercal}(n)] + E[\mathbf{x}_{I}(n)\mathbf{x}_{I}^{\intercal}(n)]$.\\
	For simplicity, we assume that the real and imaginary parts of input signal are identically distributed, therefore, imaginary and real parts of the $\mathbf{G}$ matrix are equal, i.e., $\mathbf{G_{R}} = \mathbf{G_{I}} = \mathbf{G} $ 

\end{description}
\begin{description}
	\item[$2.$] The noise sequence $\eta(n)$ is a complex zero-mean process with uncorrelated real and imaginary parts, i.e., $E[\eta(n)]$ = 0, $E[\eta_{R}(n)\eta_{I}(n)]$ = 0.
\end{description}
\begin{description}
	\item[$3.$] The input vector and noise sequence are also independent $E[\mathbf{x}(n)\eta(n)] = 0$ \cite{condition}
\end{description}
\begin{description}
	\item[$4.$] The input vector and the weight error vector are independent from each other  $E[\mathbf{x}(n) \Delta(n)]$=0
\end{description}
The weight error vector is defined as the difference between the estimated weights and the optimal weights , it is given as:
\begin{eqnarray}\label{weight-vector}
\Delta(n)=\mathbf{w}(n) - \mathbf{w}_{opt}
\end{eqnarray}
Under the assumption that the data sequence $\mathbf{x}(n)$ is statistically independent \cite{assum1,assum4,assum5}, the weight and the input vector are statistically independent.
The above suppositions are frequently used in the literature \cite{ASP}.  They are used to evaluate the performance analysis of the proposed algorithm.\\
The quantum calculus based complex form of the LMS algorithm Eq.\eqref{qLMS_final}
\begin{eqnarray}\label{complex}
\mathbf{w}(n+1)=\mathbf{w}(n)+ \mu (\mathbf{G_{R}}+\mathbf{G_{I}}) \mathbf{x}(n) e^{*}(n)
\end{eqnarray}
Now subtracting both sides of Eq.\eqref{complex} by $\mathbf{w}_{opt}$, the weight-error recursion can be given as $\Delta(n)$
\begin{eqnarray}\label{complex-weight}
\Delta(n+1)=\Delta(n)+\mu (\mathbf{G_{R}}+\mathbf{G_{I}}) \mathbf{x}(n) e^{*}(n)
\end{eqnarray}
Solving for $\Delta_{R}$, which is the real component of $\Delta(n)$ defined in Eq.\eqref{weight-vector}, substituting the value of $e_{R}(n)$ and $e_{I}(n)$ and applying the expectation results in:
\begin{eqnarray}
\begin{split}
&E[\Delta_{R}(n+1)] = \\
&E[\Delta_{R}(n)]-\mu \mathbf{G}_{R} E[\mathbf{x}_{R}(n)\mathbf{x}_{R}^{\intercal}(n)]E[\Delta_{R}(n)]\\ 
&-\mu \mathbf{G}_{R} E[\mathbf{x}_{I}(n)\mathbf{x}_{I}^{\intercal}(n)]E[\Delta_{R}(n)]
\end{split}
\end{eqnarray}
\begin{eqnarray}\label{real_mean}
E[\Delta_{R}(n+1)] = [\mathbf{I}-\mu \mathbf{G}_{R} \mathbf{R_{x}}]E[\Delta_{R}(n)] 
\end{eqnarray}

and $\mathbf{I}$ is the identity matrix.\\

$(\mathbf{I}-\mu \mathbf{G}_{R}\mathbf{D})$ is a diagonal matrix
\begin{equation} \label{lamda_ID}
\small
\mathbf{I}-\mu \mathbf{G}_{R} \mathbf{D} =
\begin{bmatrix}
1-\mu \frac{(q_{R1}+1)}{2}\lambda_{0} & & \\
& \ddots & \\
& & 1-\mu \frac{(q_{RN}+1)}{2} \lambda_{N}
\end{bmatrix} 
\end{equation}

Similarly Eq.\eqref{real_mean} can be written as:\\
\\
$E[\Delta^{'}_{R}(n+1)] =$\\
\begin{equation} \label{lamda_ID2}
\small
\begin{bmatrix}
(1-\mu \frac{(q_{R1}+1)}{2}\lambda_{0})^{n+1} & & \\
& \ddots & \\
& & (1-\mu \frac{(q_{RN}+1)}{2} \lambda_{N})^{n+1}
\end{bmatrix} E[\Delta^{'}_{R}(0)]
\end{equation}
The convergence of the $q$-CLMS algorithm depends on the diagonal elements of Eq.\eqref{lamda_ID2} and its absolute value should be less than one. 

That is to say \\
$$\displaystyle \lim_{n\to\infty} |(1-\mu \frac {q_{R}+1}{2}\lambda_{i})^{n+1}|<1,$$ for i=0 \dots N 
\begin{equation}
|(1-\mu \frac{q_{R}+1}{2}\lambda_{Rmax})|<1
\end{equation} where $\lambda_{Rmax}$ is the maximum eigenvalue of the real autocorrelation matrix. The step-size of the $q$-CLMS algorithm must be selected in the range
$$ 0 < \mu < \frac{4}{\lambda_{Rmax}(q_{R}+1)}$$

Similar methodology can be pursued for the imaginary part:
\begin{equation}\label{imag_mean}
E[\Delta_{I}(n+1)] = [\mathbf{I}-\mu \mathbf{G}_{I} \mathbf{R_{x}}]E[\Delta_{I}(n)] 
\end{equation}
$$ 0 < \mu < \frac{4}{\lambda_{Imax}(q_{I}+1)}$$

Finally, in order for the $q$-CLMS algorithm to converge, it is necessary that both the imaginary parts of its weight error must obey to the condition
\begin{equation}
0 < \mu < \frac{4}{max[\lambda_{Rmax}(q_{R}+1),\lambda_{Imax}(q_{I}+1)]}
\end{equation}

\subsection{Mean Square Analysis}\label{Sec:meansquare}
In this section, the mean square analysis of the $q$-CLMS algorithm is carried out.  To do so, we take the square of the error and its expectation gives:
\begin{eqnarray}\label{MSEE}
E[e^{2}(n)] = E[e_{o}(n) - \Delta^{H}(n)\boldsymbol{x}(n)]^{2}
\end{eqnarray}
where $H[.]$ is the Hermitian operator and $e_{o}(n)$ is the optimum output error.  Upon further simplification, Eq.\eqref{MSEE} can be given as
\begin{eqnarray}\label{diagonal1}
E[e^{2}(n)] = e^{2}_{min}  + tr[\mathbf{K}^{'}(n)\mathbf{D}]
\end{eqnarray}
where $\mathbf{K}^{'}(n)$ is the covariance of the weight error vector, $e^{2}_{min}$ is the minimum mean square error and $\mathbf{D}$ is a diagonal matrix.\\
$$\mathbf{K}^{'}(n) = \mathbf{T}^{\intercal} cov[\Delta(n)]$$
$$\mathbf{D} = diag\{\lambda_{0},\lambda_{1},\cdots,\lambda_{N}\}$$
where $\mathbf{T}$ is the unitary matrix.\\
Equation \eqref{diagonal1} can be expressed as
\begin{eqnarray}\label{mse1}
e^{2}(n) = e^{2}_{min} + \sum_{i=1}^{N} \mathbf{K}^{'}_{ii}(n) \lambda_{i}
\end{eqnarray} 
where $\lambda_{i}$ represents the eigenvalues of the input signal autocorrelation matrix and $\mathbf{K}^{'}_{ii}(n)$ shows the diagonal elements of $\mathbf{K}^{'}(n)$.  The $\lambda_{i}$ from the right side of Eq.\eqref{mse1} should be in bound to achieve a minimum error.\\
For our analysis, the $q$-CLMS algorithm can alternatively be described by:
\begin{multline}\label{transpose1}
\Delta_{R}(n+1) =\\
\Delta_{R}(n)-\mu \mathbf{G}_{R} \mathbf{x}_{R}(n)\mathbf{x}_{R}^{H}(n)\Delta_{R}(n)+\mu \mathbf{G}_{R}\mathbf{x}_{R}(n) e^{*}_{o}(n)
\end{multline}

The covariance of $\Delta_{R}^{'}(n+1)$ is given as
\begin{eqnarray}\label{40}
\mathbf{K}_{R}^{'}(n+1) = E[\Delta_{R}^{'}(n+1) \Delta_{R}^{'\intercal}(n+1)]
\end{eqnarray}
where $\mathbf{K}_{R}^{'}(n+1) = cov[\Delta_{R}^{'}(n+1)]$.\\
Taking the transpose of Eq.\eqref{transpose1}

\begin{equation}\label{41}
\Delta_{R}^{'\intercal}(n+1) =
\Delta_{R}^{'\intercal}(n)-\mu \mathbf{G}_{R} \Delta_{R}^{'\intercal}(n) \mathbf{x}_{R}^{'*}(n)\mathbf{x}_{R}^{'\intercal}(n)+\mu \mathbf{G}_{R} e^{H}_{o}(n)\mathbf{x}_{R}^{'\intercal}(n) 
\end{equation}
Multiplying Eq.\eqref{transpose1} and Eq.\eqref{41} to get Eq.\eqref{40}, we will get 9 terms, and combining all the nine terms results in
\begin{multline}
\mathbf{K}_{R}^{'}(n+1) =\mathbf{K}_{R}^{'}(n) - \mu [\mathbf{G}_{R}\mathbf{D}\mathbf{K}_{R}^{'}(n)\\+ \mathbf{G}_{R}\mathbf{K}_{R}^{'}(n)\mathbf{D}] + 2 \mu^{2}\mathbf{G}^{2}_{R}\mathbf{D}\mathbf{K}_{R}^{'}(n)\boldsymbol{D} \\+ \mu^{2}\mathbf{G}^{2}_{R}\mathbf{D}. tr[\mathbf{D} \mathbf{K}_{R}^{'}(n)] + \mu^{2}\mathbf{G}^{2}_{R} e^{2}_{min} \mathbf{D}
\end{multline}
$\mathbf{D}$ is a diagonal matrix and $e^{2}_{min}$ is the minimum mean square error.

From Eq.\eqref{mse1}, it is observable that only the diagonal elements of $\mathbf{K}_{R}^{'}(n)$ are contributing to the mean square error.
\begin{multline}\label{Eq58}
\mathbf{K}_{R}^{'}(n+1) = \mathbf{K}_{R}^{'}(n) - 2\mu \mathbf{G}_{R} \lambda_{i}\mathbf{K}_{R}^{'}(n)+2 \mu^{2} \mathbf{G}^{2}_{R}\lambda_{i}^{2}\mathbf{K}_{R}^{'}(n)\\+\mu^{2}\mathbf{G}^{2}_{R}\lambda_{i} \sum \lambda_{i}  \mathbf{K}_{R}^{'}(n)+ \mu^{2} \mathbf{G}^{2}_{R} e^{2}_{min} \lambda_{i}
\end{multline}
Considering $\lambda$ as a vector having the eigenvalues of $\mathbf{R_{x}}$ and $\mathbf{K}_{R}^{'}(n+1)$  as a vector having the diagonal elements of $cov[\Delta_{R}(n)]$, the following equation can be formed Eq.\eqref{Eq58}.
\begin{multline}\label{MSD}
\mathbf{K}_{R}^{'}(n+1) = \Bigg(I-2\mu \frac{q_{R}+1}{2}\lambda+2 \mu^{2}\Big(\frac{q_{R}+1}{2}\Big)^{2}\lambda^{2}\\+\mu^{2} \Big(\frac{q_{R}+1}{2}\Big)^{2}\lambda \lambda^{\intercal}\Bigg)\mathbf{K}_{R}^{'}(n)+\mu^{2} e^{2}_{min} \Big(\frac{q_{R}+1}{2}\Big)^{2} \lambda 
\end{multline}
\begin{equation}\label{B}
\mathbf{K}_{R}^{'}(n+1) = \mathbf{B}\mathbf{K}^{'}_{R}(n) + \mu^{2} e^{2}_{min} \Big(\frac{q_{R}+1}{2}\Big)^{2} \lambda
\end{equation}
here the elements of $\mathbf{B}$ are given by

\begin{multline}
\small
b_{ik} = \left\{ \begin{array}{rcl}
\Big(1-2\mu \frac{q_{R}+1}{2} \lambda_{i} +2 \mu^{2} \big(\frac{q_{R}+1}{2}\big)^{2} \lambda^{2}_{i} \\+ \mu^{2} \big(\frac{q_{R}+1}{2}\big)^{2} \lambda^{2}_{i}\Big)& for &i=k\\ 
2 \mu^{2} \big(\frac{q_{R}+1}{2}\big)^{2} \lambda_{i} \lambda_{k}& for & i\neq k\\ 
\end{array}\right.
\end{multline}

Consequently, a sufficient condition to guarantee convergence is to force the sum of the elements in any row
of B to be kept in the range $0 < \sum_{k=0}^{N} b_{ik} < 1$. Since 
\begin{equation}\label{diagonal}
\sum_{k=0}^{N} b_{ik} = 1-2\mu \frac{q_{R}+1}{2} \lambda_{i} +2 \mu^{2} \Big(\frac{q_{R}+1}{2}\Big)^{2} \lambda^{2}_{i} + \mu^{2} \Big(\frac{q_{R}+1}{2}\Big)^{2} \lambda_{i} \sum_{k=0}^{N} \lambda_{k}
\end{equation}

%
\begin{equation}\label{scondition}
0<1-2\mu \frac{q_{R}+1}{2} \lambda_{i} +2 \mu^{2} \Big(\frac{q_{R}+1}{2}\Big)^{2} \lambda^{2}_{i} + \mu^{2}\Big(\frac{q_{R}+1}{2}\Big)^{2} \lambda_{i} \sum_{k=0}^{N} \lambda_{k} < 1
\end{equation}

After simplification, the condition of stability is given by:
$$ 0 < \mu < \frac{2}{2 \frac{q_{R}+1}{2}\lambda_{i}+ \frac{q_{R}+1}{2}\sum_{k=0}^{N} \lambda_{k}}$$
\begin{eqnarray}\label{realS}
0 < \mu < \frac {2} {tr[\mathbf{R_{x}}\frac{q_{R}+1}{2}]}
\end{eqnarray}
In the same way, the imaginary part can be written as:
\begin{multline}
\mathbf{K}_{I}^{'}(n+1) =\mathbf{K}_{I}^{'}(n) - \mu \Big[\frac{q_{I}+1}{2}\mathbf{D}\mathbf{K}_{I}^{'}(n)+ \frac{q_{I}+1}{2}\mathbf{K}_{I}^{'}(n)\mathbf{D}\Big]\\ + 2 \mu^{2}\Big(\frac{q_{I}+1}{2}\Big)^{2}\mathbf{D}\mathbf{K}_{I}^{'}(n)\mathbf{D} + \mu^{2}\Big(\frac{q_{I}+1}{2}\Big)^{2}\boldsymbol{D}. tr[\mathbf{D} \mathbf{K}_{I}^{'}(n)] \\+ \mu^{2} \Big(\frac{q_{I}+1}{2}\Big)^{2} e^{2}_{min} \mathbf{D}
\end{multline}
\begin{eqnarray}\label{imagS}
0 < \mu < \frac {2} {tr[\mathbf{R_{x}}\frac{q_{I}+1}{2}]}
\end{eqnarray}
where Eq.\eqref{realS} and \eqref{imagS} are generally used as computation of $tr[\mathbf{R_{x}}]$  is relatively easy to estimate as compared to the  estimation of $\lambda_{max}$.

\subsection{Excess Mean Square Error (EMSE)}\label{Sec:EMSE}
The excess mean square error (EMSE) is the difference between the MSE and minimum MSE. It is considered that the minimal error is reached when the adaptive filter coefficients are frozen at their optimum values.  Although the coefficient vector converges to $\mathbf{w}_{opt}$, on average the instantaneous deviation $\Delta(n)=\mathbf{w}(n) - \mathbf{w}_{opt}$, generated by the noisy gradient estimates, causes an excess MSE.  From Eq.\eqref{MSEE} the square of error is simplified to:
\begin{eqnarray}
\xi(n)=e^{2}_{min} + E[\Delta_{R}^{\intercal}(n)\mathbf{R_{x}}\Delta_{R}(n)]
\end{eqnarray}

The excess in MSE is given by
\begin{eqnarray}
\Delta \xi_{exc}(n) \triangleq \xi(n)- e^{2}_{min} = E[\Delta_{R}^{\intercal}(n)\mathbf{R_{x}}\Delta_{R}(n)]\\  = tr\{E[\mathbf{R_{x}}\Delta_{R}(n)\Delta_{R}^{\intercal}(n)]\}
\end{eqnarray}
where $tr[\cdot]$ is the trace operator.\\
Making use of the identity that $\mathbf{TT}^{\intercal} = \mathbf{I}$, the following equation results,
\begin{eqnarray}
\Delta \xi_{exc}(n)= tr\{E[\mathbf{TT}^{\intercal}\mathbf{R_{x}TT}^{\intercal}\Delta_{R}(n)\Delta_{R}^{\intercal}(n)\mathbf{TT}^{\intercal}]\}\\  = tr\{E[\mathbf{TD} \boldsymbol{K}_{R}^{'}(n)\mathbf{T}^{\intercal}]\}
\end{eqnarray}
Therefore,
\begin{eqnarray}
\Delta \xi_{exc}(n) =  tr\{E[\mathbf{D} \mathbf{K}_{R}^{'}(n)]\}	
\end{eqnarray}
From Eq.\eqref{B}, it is possible to show that
\begin{eqnarray}
\Delta \xi_{exc}(n) = e^{2}_{min} + \sum_{i=1}^{N} \mathbf{K}^{'}_{ii}(n) \lambda_{i}
\end{eqnarray}
Since,
\begin{multline}\label{Eq74}
\mathbf{K}_{R,i}^{'}(n+1)=\\ \Bigg(1-2\mu \frac{q_{R}+1}{2} \lambda_{i}+ 2 \mu^{2}\Big(\frac{q_{R}+1}{2}\Big)^{2} \lambda_{i}^{2}\Bigg)\mathbf{K}_{R}^{'}(n)\\+\mu^{2} \Big(\frac{q_{R}+1}{2}\Big)^{2} \lambda_{i}\sum_{k=0}^{N}\lambda_{k}\mathbf{K}_{R,k}^{'}(n)+\mu^{2}e_{min}^{2} \Big(\frac{q_{R}+1}{2}\Big)^{2}  \lambda_{i}
\end{multline}
and $\mathbf{K}_{R,i}^{'}(n+1) = \mathbf{K}_{R,i}^{'}(n) $ for large $n$, we can apply a summation operation to Eq.\eqref{Eq74} in order to obtain
\begin{multline}
\sum_{k=0}^{N}\lambda_{k}\mathbf{K}_{R,k}^{'}(n) =\\
 \frac{\mu e_{min}^{2} \sum_{i=0}^{N}\frac{q_{R}+1}{2}\lambda_{i}+2\mu \sum_{i=0}^{N}\frac{q_{R}+1}{2}\lambda_{i}^{2}\mathbf{K}_{R,i}^{'}(n)}{2-\mu  \sum_{i=0}^{N}\frac{q_{R}+1}{2}\lambda_{i}}\\ \approx \frac{\mu e_{min}^{2}  \sum_{i=0}^{N}\frac{q_{R}+1}{2}\lambda_{i}}{2-\mu \sum_{i=0}^{N}\frac{q_{R}+1}{2} \lambda_{i}}\\=\frac{\mu e_{min}^{2} tr[\mathbf{R_{x}}\frac{q_{R}+1}{2}] }{2-\mu  tr[\mathbf{R_{x}}\frac{q_{R}+1}{2}]}
\end{multline}
where the term $2\mu\sum_{i=0}^{N}\frac{q_{R}+1}{2}\lambda_{i}^{2}\mathbf{K}_{R,i}^{'}(n)$ 
was considered quite insignificant in comparison to the remaining terms in the numerator. This assumption is difficult to justify, but it holds true for small values of $\mu$.
The excess mean-square error can then be expressed as
\begin{equation}\label{EMSE1}
\Delta \xi_{exc}(n) = \lim_{i\to\infty}\frac{\mu \sigma^{2} tr[\mathbf{R_{x}}\frac{q_{R}+1}{2}] }{2-\mu  tr[\mathbf{R_{x}}\frac{q_{R}+1}{2}]}
\end{equation}
Similarly for imaginary part,
\begin{equation}
\Delta \xi_{exc}(n) = \lim_{i\to\infty}\frac{\mu \sigma^{2} tr[\mathbf{R_{x}}\frac{q_{I}+1}{2}] }{2-\mu  tr[\mathbf{R_{x}}\frac{q_{I}+1}{2}]}
\end{equation}
In the case where there is a sufficient order of the adaptive filter to model the operation that generated $d(n)$, the minimum MSE attained is equivalent to the variance of the noise added, given by $\sigma^{2}$ .\\

\subsection{Mean Square Deviation (MSD)}\label{Sec:MSD}
The mean square deviation (MSD	) is defined as \\$MSD(n) = tr\{E[\mathbf{K}_{R}^{'}(n+1)]\}$, consequently 
\begin{multline}
\small
MSD(n+1) = \Bigg(I-2\mu \frac{q+1}{2} \lambda+2 \mu^{2}\Big(\frac{q+1}{2}\Big)^{2}\lambda^{2}\\+\mu^{2}\Big(\frac{q+1}{2}\Big)^{2}\lambda \lambda^{\intercal}\Bigg)MSD(n)+\mu^{2} e^{2}_{min}\Big(\frac{q+1}{2}\Big)^{2} \lambda 
\end{multline}
\subsection{Misadjustment}
The ratio between the excess MSE ($\xi_{exc}$) and the minimum MSE ($\xi_{min}$) is known as the Misadjustment $M$ , the misadjustment is given by
\begin{equation}
M \triangleq \frac{\Delta \xi_{exc}}{e_{min}^{2}} \approx \frac{\mu tr[\mathbf{R_{x}}\frac{q+1}{2}] }{2-\mu tr[\mathbf{R_{x}}\frac{q+1}{2}]}
\end{equation}
The analytical results for MSE, EMSE and Misadjustment are summarized in Table.\ref{analytical}.

\begin{table}[!ht]
\centering
\begin{tabular}{|l|l|l|}

\hline
\multirow{2}{*}{}      & \multicolumn{2}{c|}{\textit{\textbf{Analytical Results}}}                                                                                                                                                           \\ \cline{2-3} 
                       & \multicolumn{1}{c|}{\textbf{Real}}                                                                       & \multicolumn{1}{c|}{\textbf{Imaginary}}                                                                  \\ \hline
\textbf{MSE}           & $0 < \mu < \frac {2} {tr[\mathbf{R_{x}}\frac{q_{R}+1}{2}]}$                                              & $0 < \mu < \frac {2} {tr[\mathbf{R_{x}}\frac{q_{I}+1}{2}]}$                                              \\ \hline
\textbf{EMSE}          & $\frac{\mu \sigma^{2} tr[\mathbf{R_{x}}\frac{q_{R}+1}{2}] }{2-\mu  tr[\mathbf{R_{x}}\frac{q_{R}+1}{2}]}$ & $\frac{\mu \sigma^{2} tr[\mathbf{R_{x}}\frac{q_{I}+1}{2}] }{2-\mu  tr[\mathbf{R_{x}}\frac{q_{I}+1}{2}]}$ \\ \hline
\textbf{Misadjustment} & $ \frac{\mu tr[\mathbf{R_{x}}\frac{q_{R}+1}{2}] }{2-\mu tr[\mathbf{R_{x}}\frac{q_{R}+1}{2}]}$            & $\frac{\mu tr[\mathbf{R_{x}}\frac{q_{I}+1}{2}] }{2-\mu tr[\mathbf{R_{x}}\frac{q_{I}+1}{2}]}$             \\ \hline
\end{tabular}
\caption{Analytical Results}
\label{analytical}
\end{table}
\section{The proposed Enhanced \emph{q}-CLMS (E\emph{q}-CLMS)}\label{Sec:time-varying}
The selection of the q-parameter shows greater importance in the performance of $q$-calculus based algorithms.  It can be analyzed in sub-section \ref{Sec:sensitivity} of the experimental section.  Typically this value is manually chosen.  A larger value results in a rapid convergence but yields a larger steady-state error.  On the other hand, smaller values of $q$ slow down the convergence while reducing the steady-state error.  A trade-off is therefore required by manually choosing a suitable value of the $q$-parameter.  

To automate this choice, we modify the proposed $q$-CLMS algorithm by making the $q$-parameter time varying: 
\begin{equation}\label{EqCLMS_final}
\mathbf{w}(n+1) = \mathbf{w}(n) + \mu \mathbf{x}(n)e^{*}(n)\odot\mathbf{q}(n)
\end{equation}
where $\mathbf{q}(n)$ is the time-varying parameter, which can be calculated by taking the error-correlation energy \cite{EqLMS}.
The proposed algorithm takes large steps for larger error, and decreases the size of the number of steps with a lower error.  Highlighting the fact that, the proposed method is absolutely automatic and does not require any other tuning factors as compared to the time-varying $q$-LMS.  The proposed update rule for the $\emph{q}$-parameter is described below:\\
\begin{itemize}
	\item \textbf{\textit{Step1}}: Initialize by taking any positive random values in vector $\mathbf{q}$ .
	\item \textbf{\textit{Step2}}: To update the first entry $q_1$ of the vector $\mathbf{q}$,  use the instantaneous error related to the weight of the instant input tap, i.e., 
	\begin{equation}
	q_{1}(n+1) = \frac{1}{M+1}\{|e(n)|+\sum_{k=1}^{M}\mathbf{q}_{k}(n)\},
	\end{equation}
	The length of the filter is given by the symbol $M$.
	\item \textbf{\textit{Step3}}:  In order to maintain fast convergence rate and to avoid divergence, the following conditions will be evaluated:
	\begin{equation}\label{q_parameter}
	\mathbf{q}(n+1) = \left\{ \begin{array}{rcl}
	\frac {1} {\lambda_{max}} & if & |q_{1}(n+1)| > \frac {1} {\lambda_{max}},\\ 
	q_{1}(n+1) & otherwise,\\ 
	\end{array}\right.
	\end{equation}
	where $\lambda_{max}$ is the maximum eigenvalue of the input auto-correlation matrix.  
	\item \textbf{\textit{Step4}}: Except for the first entry, update all entries of the vector $\mathbf{q}$, simply by shifting:    
	\begin{equation}\label{qn}
	q_{l+1}(n+1) = q_{l}(n),
	\end{equation}
	where $1 < l < M-1$
	\item \textbf{\textit{Step5}}: Repeat steps 2 to 5, for next iterations, 
\end{itemize}

Conclusively, the equation for weight-update of the proposed $Eq$-CLMS algorithm can be given as:
\begin{equation}\label{EqLMS_final}
\mathbf{w}(n+1) =\mathbf{w}(n) + \mu e^{*}(n)\mathbf{x}(n)\odot\mathbf{q}(n),
\end{equation}
where $\odot$ indicates the element-wise multiplication.
\section{The proposed \emph{q}-CLMS algorithm as a whitening filter}\label{Sec:whitening}
Due to the spread of eigen values of the input correlation matrix, the convergence speed of LMS algorithm is limited as it shows dependency on eigen values.  Explicitly, the time constant $\tau_{a}$ is given as \cite{ASP}
\begin{equation}\label{18}
\frac{-1}{ln(1-\mu \lambda_{max})} \leqslant \tau_{a} \leqslant \frac{1}{ln(1-\mu \lambda_{min})}
\end{equation}
where $ln$ shows the logarithmic function, and the minimum and maximum eigenvalues of the input correlation matrix are represented by $\lambda_{min}$ and $\lambda_{max}$, respectively.\\
By taking the above observation into consideration, we established another application of the $q$-gradient.  Explicitly, for a faster convergence rate of the LMS algorithm, the $q$ factor can be selected in such a way as to make the LMS filter behaves as a whitening filter (shown in Fig. \ref{Wfilter}). 
For evaluation purposes, the transient analyses of the weight error vector of the $q$-CLMS given in Eqs.\eqref{real_mean} and \eqref{imag_mean} is studied.  Note that the time constant related with the $m^{th}$ mean weight error tap $\Delta_{R,m}(n)$ is given as
\begin{equation}
\tau_{m} = \frac{-1}{ln(1-\frac{\mu (q_{m}+1) \lambda_{m}}{2})},   1\leqslant m \leqslant M
\end{equation}
Thus selecting the $q_{m}$ parameter such that
\begin{equation}\label{20}
\frac{(q_{m}+1)}{2} = \frac{1}{\lambda_{m}} \quad \textrm{or} \quad q_{m} = \frac{2-\lambda_{m}}{\lambda_{m}},  1\leqslant m \leqslant M
\end{equation}
the time constant $\tau_{m}$ reduces to:
\begin{equation}
\tau_{m} = 	\frac{-1}{ln(1-\mu )},   1\leqslant m \leqslant M 
\end{equation}
Essentially, by taking the $q$ values as shown in  Eq.\eqref{20} or by taking $\mathbf{G}$  = $\mathbf{R_{x}}^{-1}$, the $q$-CLMS method vanishes the response of the input correlation matrix.  As a result, this will enhance the speed of the proposed method.  Eventually, the condition for stability is given by:
\begin{equation}
0 < \mu < 2\\
\end{equation}

\subsection{Computational Complexity Analysis}

Table \ref{tab:complexity} shows a comparison of the computational complexity of different variants of CLMS in terms of the number of operations required for the adaptation process.

\begin{table}[!ht]
\centering
\begin{tabular}{|c|c|c|c|}
\hline
\multirow{2}{*}{\textbf{Method}} &
  \multicolumn{2}{c|}{\textbf{Operations}} &
  \multirow{2}{*}{\textbf{Big O}} \\ \cline{2-3} 
              & x    & +    &   \\ \hline
\textbf{CLMS}     & 2M+1 & 2M  & M  \\ \hline
\textbf{NCLMS}    & 3M+2 & 3M+1 &  M \\ \hline
\textbf{CFLMS \cite{Shah2017}}    & 4M+4 & 6M & M \\ \hline
\textbf{FoCLMS \cite{Ahmad2021}}   & 3M+3 & 5M &  M \\ \hline
\textbf{\emph{q}-CLMS}  & \textbf{3M+1} & \textbf{2M}  &  \textbf{M} \\ \hline
\textbf{E\emph{q}-CLMS}  & \textbf{2M+2} & \textbf{3M+1}  &  \textbf{M} \\ \hline
\end{tabular}
\caption{Computational complexities of different algorithms in terms of the number of unknown weights (M)}
\label{tab:complexity}
\end{table}

Let M be the number of unknown filter weights. The CLMS method takes $2M+1$ multiplications and $2M$ additions, while the NCLMS algorithm takes $3M+2$ multiplications and $2M$ additions. Similarly, the CFLMS and FoCLMS, which are fractional variants of LMS, need $4M+4$ and $3M+3$ multiplications and $6M$ and $5M$ additions respectively. Interestingly the proposed quantum calculus based variants of CLMS algorithms, i.e., \emph{q}-CLMS and E\emph{q}-CLMS, require $3M+1$ and $2M+2$ multiplications, and $2M$ and $3M+1$ additions respectively. The proposed E\emph{q}-CLMS and \emph{q}-CLMS require fewer multiplications and additions compared to the NCLMS and fractional variants of the CLMS.  In particular, the \emph{q}-CLMS and E\emph{q}-CLMS require $1$ unit and $M$ times fewer multiplications and $M+1$ times fewer additions compared to the NCLMS.   Similarly, the \emph{q}-CLMS and E\emph{q}-CLMS require $M+3$ and $2M+2$ fewer multiplications and $4M$ and $3M$ fewer additions compared to the CFLMS.  Moreover, FoCLMS requires $2$ units and $M+1$ more multiplications as well as $3M$ and $2M$ times more additions than the proposed \emph{q}-CLMS and E\emph{q}-CLMS algorithm.  In an overall sense, the proposed \emph{q}CLMS and E-\emph{q}CLMS are shown to be computationally efficient compared to the state-of-the-art relevant approaches.

	\section{Experimental Setup and Results}\label{Sec:Sim}
\begin{figure}[!ht]
	\begin{center}
		\centering
		\includegraphics[width=0.8\textwidth]{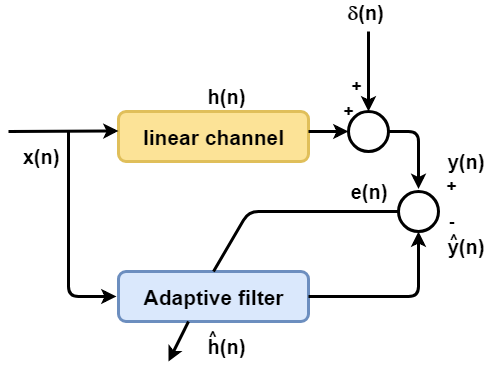}
	\end{center}
	\caption{Adaptive learning algorithm based channel estimation.}
	\label{model}
\end{figure}

The performance of the $\emph{q}$-CLMS algorithm is examined in this section for a system identification task.  Consider channel estimation shown in Fig. \ref{model}.

\begin{equation}\label{math_model}
y(n) =
h_{1}x(n)+h_{2}x(n-1)+h_{3}x(n-2)+h_{4}x(n-3)+h_{5}x(n-4)
\end{equation}
The mathematical model of the system is shown in Eq.\eqref{math_model}, it is the FIR filter representation for the adaptive filter block (refer to Fig \ref{model}).  Where input and output of the system expressed as $\mathbf{x}(n)$ and $y(n)$, respectively.  The white Gaussian noise is considered as the disturbance for the system in this case.  For experimentation purpose, $\mathbf{x}(n)$ consists of $1\times10^{3}$ complex randomly generated samples taken from the Gaussian distribution with zero mean and unity variance.  The impulse response of the system is given by $\mathbf{h}(n)$ in Eq.\eqref{math_model} while $e(n)$, $\mathbf{\hat{h}}(n)$ and $\hat{y}(n)$ are the estimated error, impulse response and output of the system, respectively.  For the performance assessment, the actual and estimated weights are compared by means of the Normalized Weight Deviation (NWD).  Explicitly, we define \begin{equation}\label{NWD}
\mbox{NWD}=\frac{\left\Vert \bf h-\bf w \right\Vert}{\left\Vert\bf h\right\Vert}
\end{equation} 
where estimated weight-vector is denoted by $\mathbf{w}$ while the actual impulse response of the channel is given by $\mathbf{h}$.  For all the simulations, the adaptive filter which is used has the same length as that of the unknown system.  The objectives of the simulation are as follows:
\begin{description}
	\item[$\bullet$] To observe the sensitivity of the proposed method corresponding to the $q$ parameter.
	
	\item[$\bullet$] To validate the derived analytical results for both the steady state and transient analysis.

	\item[$\bullet$] To investigate the behavior of the proposed E$q$-CLMS algorithm and compare it with the benchmark algorithms i.e., CLMS, NCLMS and VSS-CLMS.

	\item[$\bullet$]  To evaluate the performance of the proposed E$q$-CLMS method for correlated imaginary and real components of the system.
	\item[$\bullet$] To investigate the behavior of $q$-CLMS as a whitening filter.

	\item[$\bullet$] To compare the performance of the proposed E\emph{q}-CLMS algorithm with the contemporary approaches i.e., CFLMS and FoCLMS for the channel equalization problem.
\end{description}

\subsection{Sensitivity analysis of the \emph{q}-CLMS algorithm}\label{Sec:sensitivity}
\begin{figure}[!ht]
	\begin{center}
		\centering
		\includegraphics[width = 0.8\textwidth]{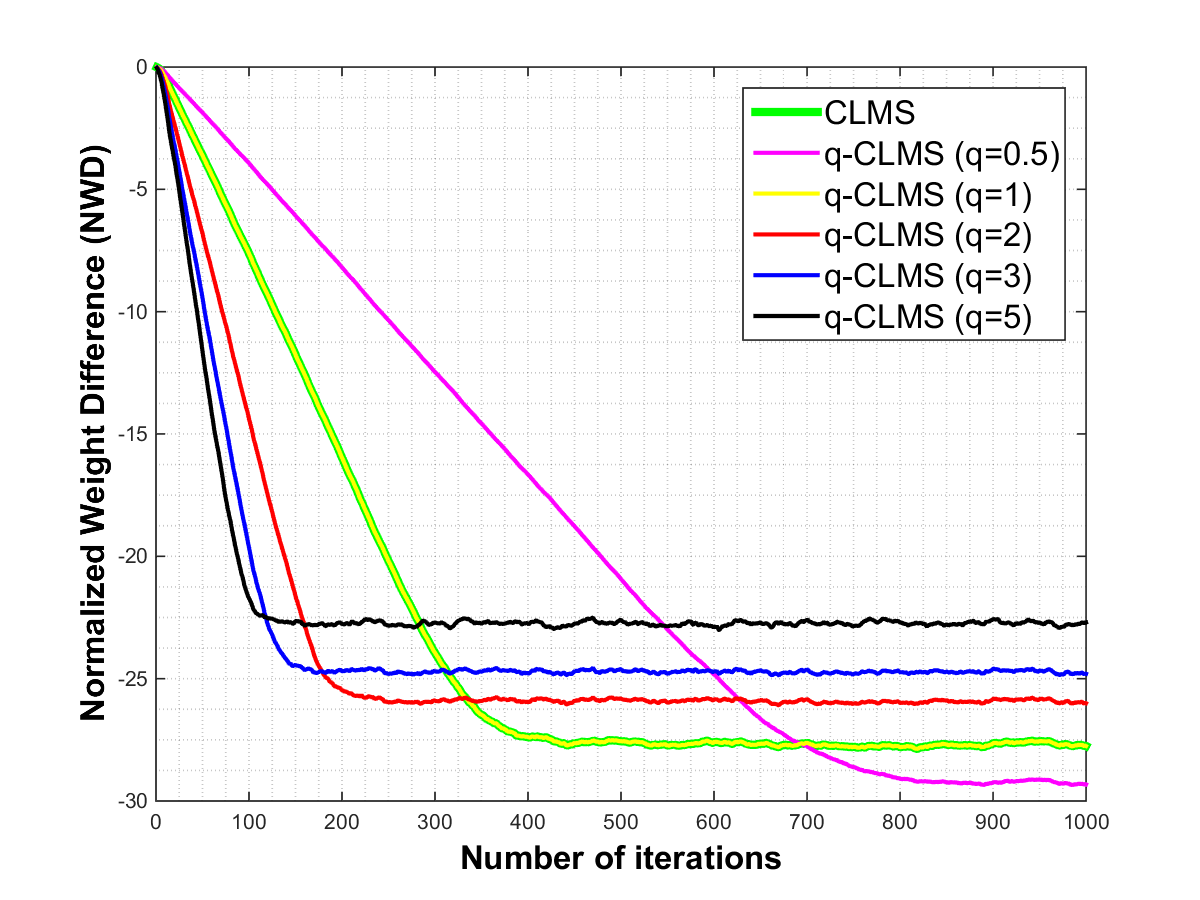}
	\end{center}
	\caption{NWD behaviour for the proposed $q$-CLMS and the conventional CLMS algorithm.}
	\label{sensitivity}
\end{figure}

In this experiment, we observe the sensitivity of the $q$-CLMS algorithm to the $q$ parameter.  In particular, we choose a system identification task and study the comparison between the Normalized Weight Difference (NWD) learning curves of the proposed $q$-CLMS algorithm with various $q$ values and compare the results with those obtained via conventional CLMS algorithm (refer to Fig.\ref{sensitivity}).  The disturbance which is introduced in the system as a noise  is a zero mean i.i.d. sequence with variance 0.01.  For convergence the proposed approach took 1000 number of iterations and the mean values are taken for 100 independent runs.  We evaluated five different values of $q$ which are $0.5$ , $1$, $2$, $4$, and $8$ for the $q$-CLMS algorithm . Figure \ref{sensitivity} clearly shows that for $q=1$ the learning curve exactly matches with the conventional CLMS method showing that for $q=1$ the algorithm shows the conventional CLMS case.  It is observable that for greater values of $q$, the proposed $q$-CLMS algorithm exhibits a faster convergence accompanied with a larger steady-state error.  Whereas for smaller values of $q$, a comparatively smaller steady-state error is reported but with a slower convergence.

\subsection{Validation of analytical results of the proposed \emph{q}-CLMS algorithm}\label{valid}
\begin{figure}[!ht]
	\begin{center}
		\centering
		\includegraphics[width=0.8\textwidth]{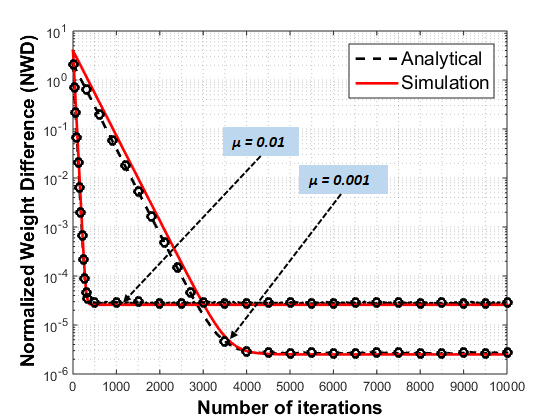}
	\end{center}
	\caption{Simulation and analytical MSE behavior of the $q$-CLMS algorithm: For $\mu = 0.01$ and $\mu = 0.001$.}
	\label{meansquare}
\end{figure}

\begin{figure}[!ht]
	\begin{center}
		\centering
		\includegraphics[width=0.8\textwidth]{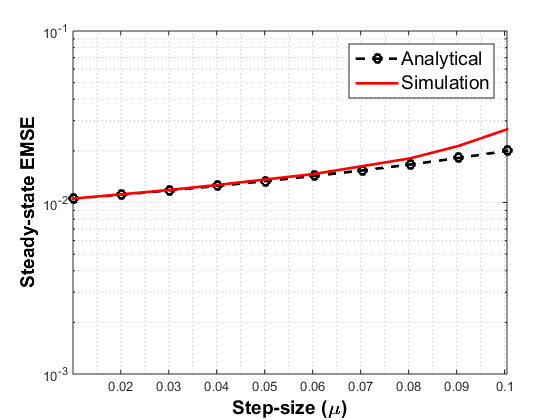}
	\end{center}
	\caption{Simulation and analytical EMSE behavior of the $q$-CLMS algorithm: Excess MSE gradually increase with increase in step-size.}
	\label{EMSE}
\end{figure}

To validate our theoretical findings, we compare the the derived analytical results with the simulation results.   The $q$-CLMS algorithm's transient and steady-state performances are essentially examined. For two choices of the step-size i.e., 0.01 and 0.001, experimental MSE learning curves are compared to those generated using the derived expression in Eq.\eqref{MSD}. The algorithm took 10000 iterations to reach convergence, and the mean value was obtained from 1000 independent runs. For both large and small step-size values, Fig.\ref{meansquare} indicates an outstanding match between theory and simulation. The step-size values are compared to the simulation findings in the second experiment shown in Fig.\ref{EMSE}, which are plotted against the analytical values of the steady-state EMSE determined in Eq.\eqref{EMSE1}. The relationship between the theoretical and simulation results is also shown in Fig. \ref{EMSE}.  

\subsection{Performance of the proposed E\emph{q}-CLMS algorithm} \label{comp}
\begin{figure}[!ht]
	\begin{center}
		\centering
		\includegraphics[width=0.8\textwidth]{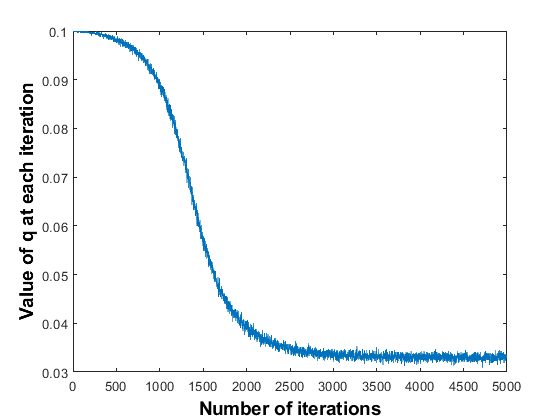}
	\end{center}
	\caption{The value of \emph{q} after each iteration when operating on the SNR value of 30 dB.}
	\label{qvalue}
\end{figure}

\begin{figure}[!ht]
\centering
\begin{minipage}[b]{.49\textwidth}
\includegraphics[width=\textwidth]{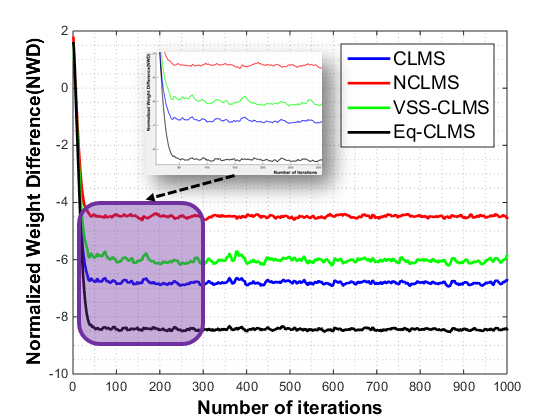}
\caption*{(a)}
\end{minipage}
\begin{minipage}[b]{.49\textwidth}
\includegraphics[width=\textwidth]{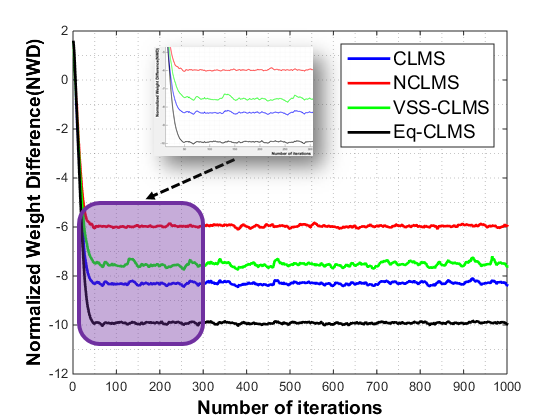}
\caption*{(b)}
\end{minipage}
\begin{minipage}[b]{.49\textwidth}
\includegraphics[width=\textwidth]{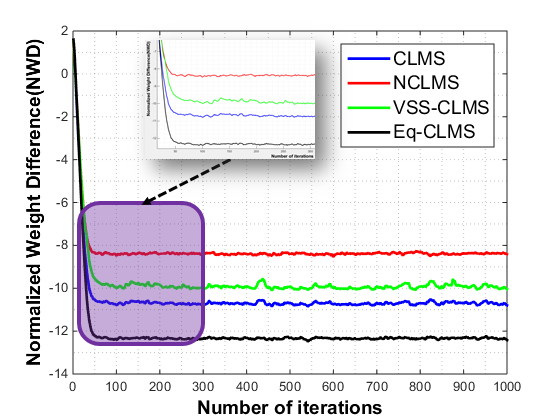}
\caption*{(c)}
\end{minipage}
\begin{minipage}[b]{.49\textwidth}
\includegraphics[width=\textwidth]{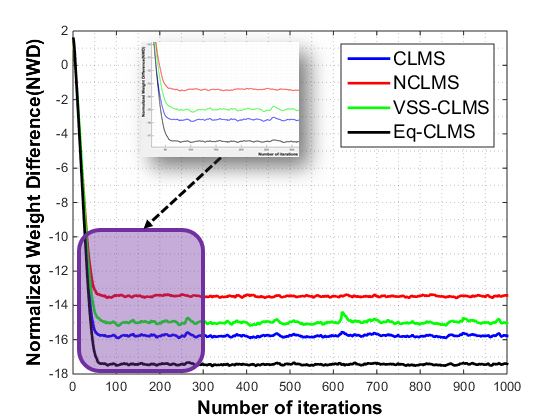}
\caption*{(d)}
\end{minipage}
\begin{minipage}[b]{.52\textwidth}
\includegraphics[width=\textwidth]{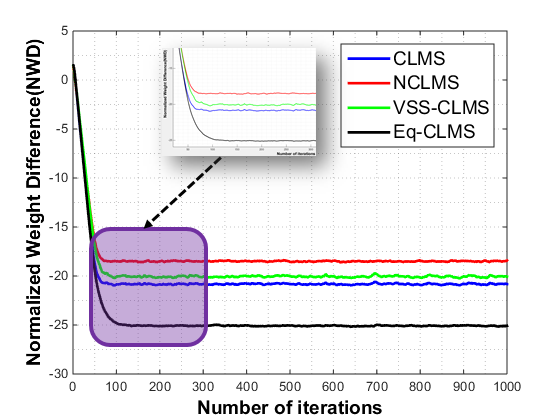}
\caption*{(e)}
\end{minipage}
\caption{NWD curves for the CLMS, NCLMS, VSS-CLMS, and the proposed $Eq$-CLMS algorithm with the SNR of (a) $2$ dB, (b) $5$ dB,(c) $10$ dB, (d) $20$ dB, and (e) $30$ dB.}
\label{SNRfigure}
\end{figure}

In this set of experiments, we evaluate the performance of the E$q$-CLMS algorithm as derived in Section \ref{Sec:time-varying}.  Essentially the self-regulating technique, proposed in Eqs.\eqref{q_parameter} and \eqref{qn}, provides an automatic adjustment. The experiment is implemented with five different SNR levels of $2$ dB, $5$ dB, $10$ dB. $20$ dB and $30$ dB.  
Note that the proposed E$q$-CLMS algorithm is completely automatic and does not require a hard-coded value of the $q$ parameters.  By virtue of the recursions proposed in Eqs.\eqref{q_parameter} and \eqref{qn}, the entries of $\mathbf{q}$ are automatically regularized to attain the best performance.  For our experiments, an initialization of $\mathbf{q}=[1 \ 1 \ 1 \ 1 \ 1]$ is adopted.  \\
Figure \ref{SNRfigure} shows the performance of the proposed E\emph{q}-CLMS algorithm when operating on five different noise levels. The proposed E\emph{q}-CLMS algorithm obtained the lowest steady-state error of $-25.13$ dB when operating on the highest value of SNR i.e., $30$ dB, while the contemporary approaches CLMS, NCLMS and VSS-CLMS achieved higher steady-state values of $-20.82$ dB, $-18.5$ dB and $-20.03$ dB, respectively.  For the lowest SNR which is $2$ dB, the proposed E\emph{q}-CLMS algorithm outperformed the conventional CLMS and VSS-CLMS algorithm by a margin of $2$ dB and NCLMS by a margin of $3.938$ dB.  Similarly for $5$, $10$ and $20$ dB the lowest steady-state error is achieved by the proposed E\emph{q}-CLMS algorithm i.e, $-9.912$ dB, $-12.37$ dB and $-17.39$ dB respectively.  Steady-state error attained for five  different noise levels for CLMS, NCLMS, VSS-CLMS and E\emph{q}-CLMS algorithms are summarized in table \ref{SNR}.  To show the adaptivity of \emph{q} parameter, in Figure \ref{qvalue} for 30 dB noise level we showed the values of \emph{q} for each update iteration. 


\begin{table}[!ht]
\centering
\small
\caption{Comparison of steady-state performance}
\label{SNR}
\begin{tabular}{|l|c|c|c|c|c|}
\hline
\multicolumn{1}{|c|}{\multirow{2}{*}{\textit{\textbf{Algorithms}}}} & \multicolumn{5}{c|}{\textit{\textbf{\begin{tabular}[c]{@{}c@{}}SNR Values\\ (dB)\end{tabular}}}} \\ \cline{2-6} 
\multicolumn{1}{|c|}{}                                              & \textbf{2}        & \textbf{5}        & \textbf{10}       & \textbf{20}       & \textbf{30}      \\ \hline
\textit{\textbf{CLMS}}                                              & -6.87             & -8.32             & -10.75            & -15.79            & -20.82           \\ \hline
\textit{\textbf{NCLMS}}                                             & -4.521            & -5.929            & -8.39             & -13.92            & -18.5            \\ \hline
\textit{\textbf{VSS-CLMS}}                                          & -6.07             & -7.48             & -9.949            & -15.01            & -20.03           \\ \hline
\textit{\textbf{Eq-CLMS}}                                           & \textbf{-8.459}   & \textbf{-9.912}   & \textbf{-12.37}   & \textbf{-17.39}   & \textbf{-25.13}  \\ \hline
\end{tabular}
\end{table}

\newpage

\subsection{Complex signal with correlated real and imaginary data}
In this experiment, we evaluate the performance of the proposed algorithm for complex data with correlated imaginary and real parts. The complex input signal $\mathbf{x}(n)$  of size $1 \times 5000$ was taken from the non-Gaussian distribution with correlated imaginary and real parts and impulse response of the system $\mathbf{h}(n)$ was taken to be complex randomly generated samples of size $1 \times 5$. Experiments were performed for both \emph{q}-CLMS and E\emph{q}-CLMS algorithms, simulations were run for $5000$ iterations and the mean results were computed for 1000 independent runs.\\ \\
To evaluate the proposed \emph{q}-CLMS algorithm, a fixed value for the G matrix i.e, $\mathbf{q}=[3 \ 3 \ 3 \ 3 \ 3]$ is taken. The proposed \emph{q}-CLMS algorithm attained the lowest steady-state error of $-23.28$ dB whereas the conventional CLMS algorithm attained a larger value of $-21.44$ dB.
\begin{figure}[!ht]
	\begin{center}
		\centering
		\includegraphics[width=0.8\textwidth]{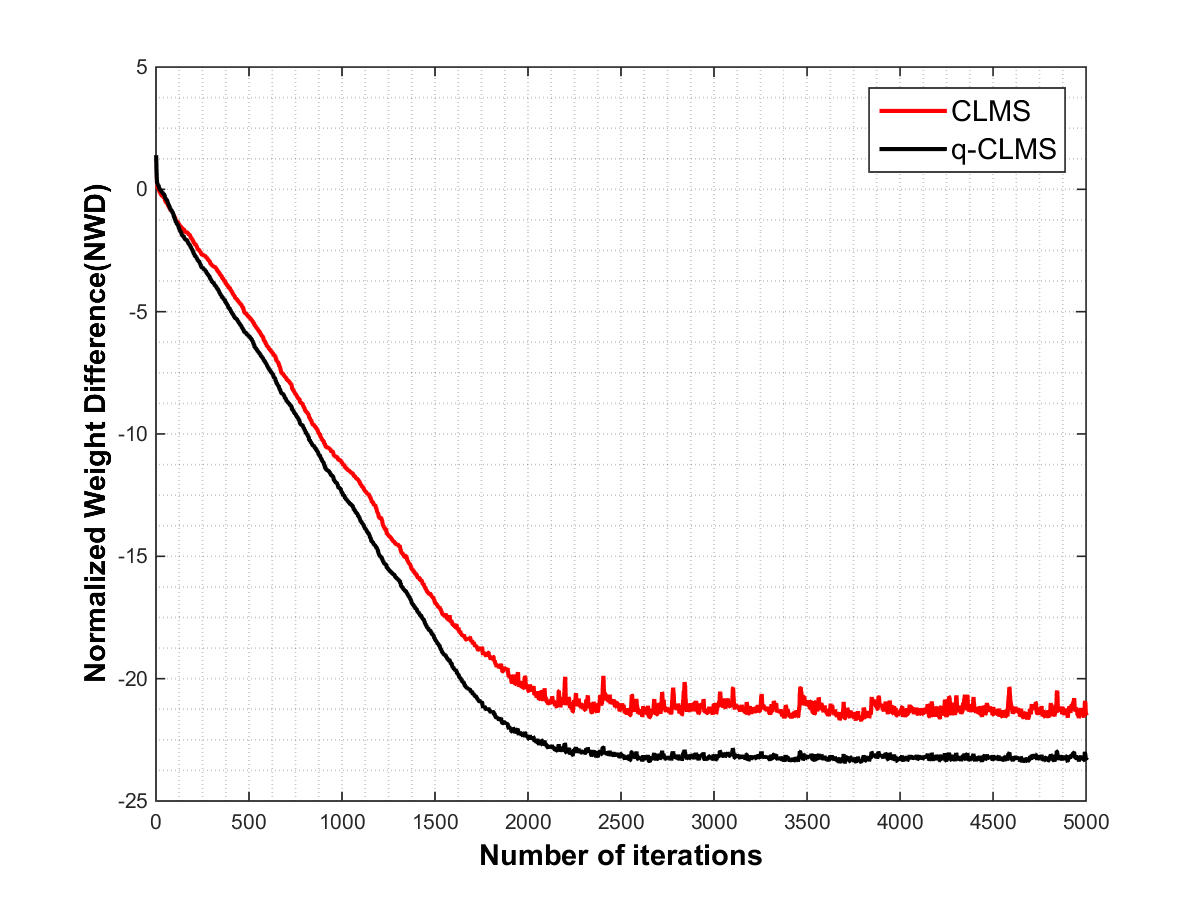}
	\end{center}
	\caption{Performance curve for data having correlated real and imaginary parts with a fixed $\mathbf{G}$ matrix.}
	\label{CRIG}
\end{figure}

The second set of experiments was conducted using an adaptive 'q' parameter, an initial value of $\mathbf{q}=[0.02 \ 0.02 \ 0.02 \ 0.02 \ 0.02]$ was adopted.  The proposed E\emph{q}-CLMS method attained the least steady-state error of $-25$ dB whereas the conventional CLMS algorithm attained a larger value of $-20.8$ dB.
\begin{figure}[!ht]
	\begin{center}
		\centering
		\includegraphics[width=0.8\textwidth]{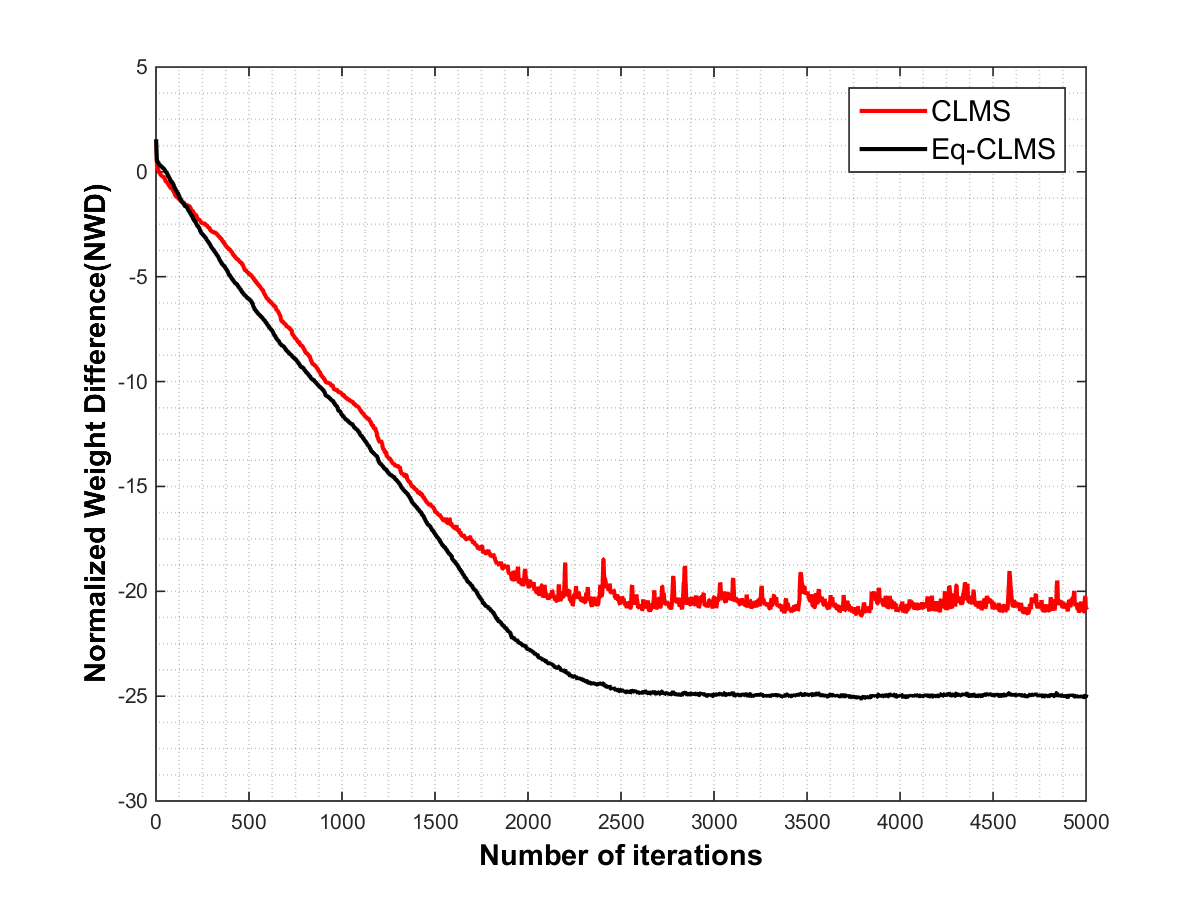}
	\end{center}
	\caption{Performance curve for correlated complex signal with varying \emph{q}-parameter.}
	\label{CRI}
\end{figure}
\subsection{Whitening behavior of the proposed \emph{q}-CLMS algorithm}
\begin{figure}[!ht]
	\begin{center}
		\centering
		\includegraphics[width=0.8\textwidth]{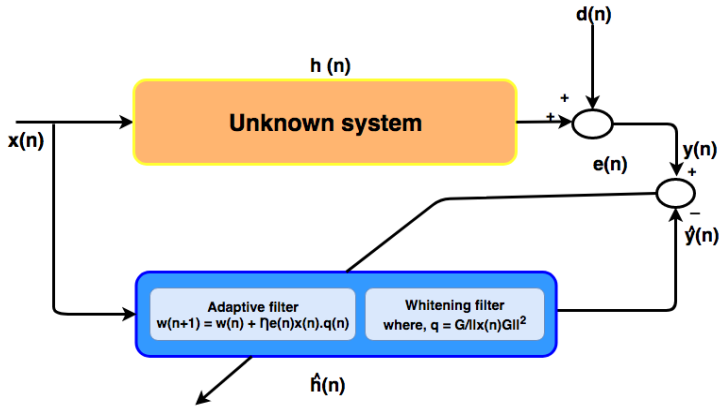}
	\end{center}
	\caption{System identification and whitening filter.}
	\label{Wfilter}
\end{figure}
\begin{figure}[!ht]
	\begin{center}
		\centering
		\includegraphics[width=0.8\textwidth]{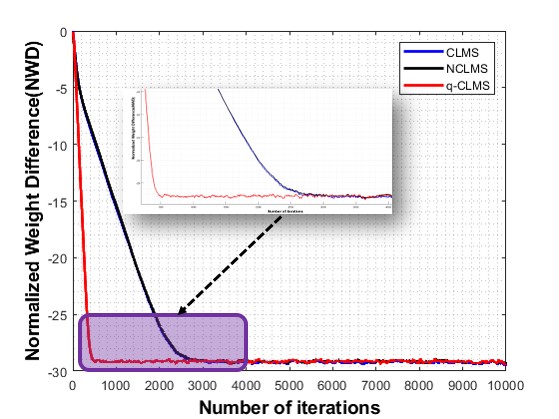}
	\end{center}
	\caption{Comparison of the NWD behaviour of the whitening $q$-CLMS and the conventional CLMS algorithm: The proposed approach took $1750$ less number of iterations compared to CLMS and NCLMS.}
	\label{whitening_1}
\end{figure}

This section examines the whitening behavior of the $q$-CLMS algorithm.  A graphical model of the system identification with adaptive whitening filter is shown in Figure \ref{Wfilter}.  As mentioned in section \ref{Sec:whitening}, the q-CLMS algorithm vanished the effect of the input correlation.  A non-Gaussian input signal is considered to evaluate the whitening behavior of the proposed $q$-CLMS algorithm.  For convergence the algorithm took 10000 iterations and the mean value is taken for 100 independent runs.  The NWD learning curve of the \emph{q}-CLMS method ($\mathbf{G}$ = $\mathbf{R_{x}}^{-1}$) is compared to those of the conventional CLMS and NCLMS methods in Fig.\ref{whitening_1}.  From the results, it can be seen that the CLMS and the NCLMS algorithms achieve equal steady-state error of $-29.26$ dB in $2600$ iterations, approximately.  The proposed q-CLMS (as a whitening filter) comprehensively outperforms the two competitors, achieving the same steady state error of $-29.26$ dB in $500$ iterations only which is $5.2 \times$ faster than the two competitors.  The use of Eq.\eqref{20} to select the $q$ parameter allows the $q$-CLMS method to converge regardless of the input correlation, as explained in Section \ref{Sec:whitening}. Furthermore, the results in Figure \ref{whitening_1} supported the claim for two extreme values of the correlation factor, $\sigma xx$ = $0.99$ and $\sigma xx$ = $0.1$, the $q$-CLMS method with whitening $q$ selection is simulated. The whitening $q$ selection ensures that the $q$-CLMS algorithm's convergence is independent of the input correlation, as shown in Figure \ref{whitening_1}.  

\begin{figure}[!ht]
	\begin{center}
		\centering
		\includegraphics[width=0.8\textwidth]{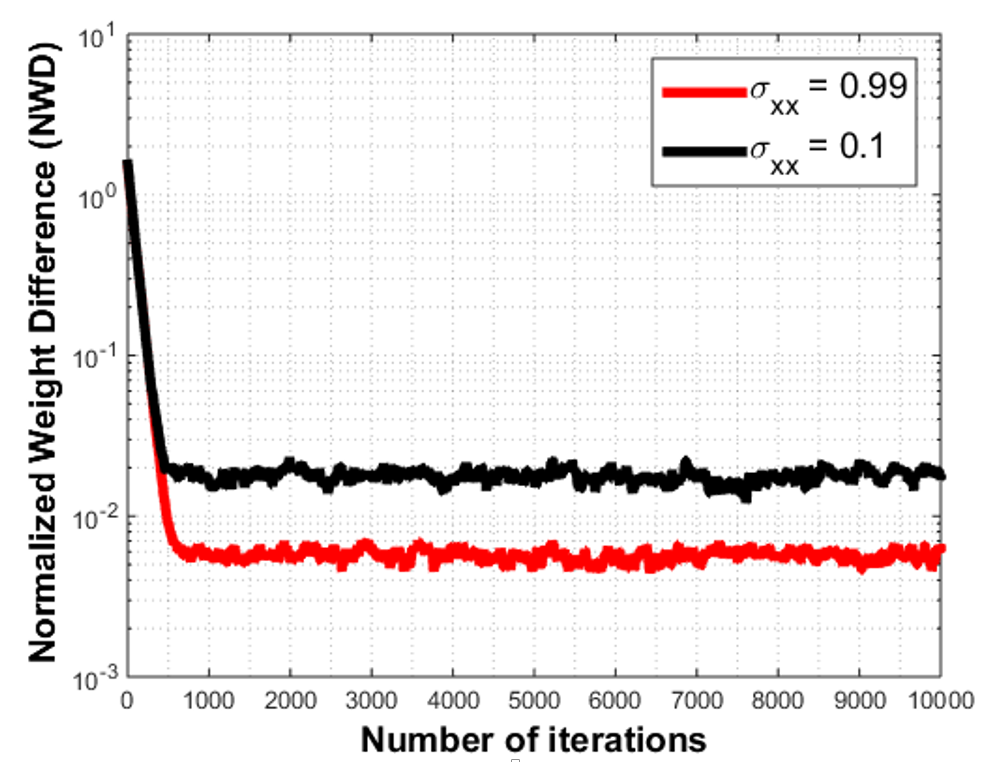}
	\end{center}
	\caption{NWD behaviour of the whitening $q$-CLMS for two extreme values of input variances i.e., $\sigma_{xx} = 0.99$ and $\sigma_{xx} = 0.1$. Note that the rate of convergence of the \emph{q}-CLMS is same for both input distributions.}
	\label{whitening}
\end{figure}

\subsection{Comparative analysis of the proposed E\emph{q}-CLMS algorithm for the problem of channel equalization}

\begin{figure}[!ht]
	\begin{center}
		\centering
		\includegraphics[width=0.8\textwidth]{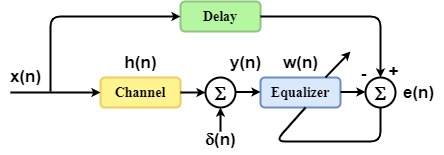}
	\end{center}
	\caption{Adaptive channel equalization.}
	\label{Wchannel}
\end{figure}

\begin{table}[!ht]
\centering
\begin{tabular}{|l|c|c|}
\hline
\multicolumn{1}{|c|}{\multirow{2}{*}{\textbf{Algorithm}}} &
  \multicolumn{2}{c|}{\textbf{\begin{tabular}[c]{@{}c@{}}Convergence point \\ (number of iterations x 5000)\end{tabular}}} \\ \cline{2-3} 
\multicolumn{1}{|c|}{} & \textbf{30 dB SNR} & \textbf{40 dB SNR} \\ \hline
CLMS                   & 103                & 110                \\ \hline
CFLMS (0.9)            & 110               & 111               \\ \hline
CFLMS (0.75)           & 121               & 127               \\ \hline
FoCLMS (0.9)           & 100                & 100                \\ \hline
FoCLMS (0.75)          & 149                & 150                \\ \hline
\textbf{Eq-CLMS}       & \textbf{51}        & \textbf{52}        \\ \hline
\end{tabular}
\caption{The Convergence point of various approaches including CLMS, CFLMS, FoCLMS (for two distinct values of fractional power i.e., 0.75 and 0.9) and the proposed Eq-CLMS algorithm.}
\label{tab:resultsexp}
\end{table}

\begin{figure}[!ht]
\centering
\begin{minipage}[b]{.7\textwidth}
\includegraphics[width=\textwidth]{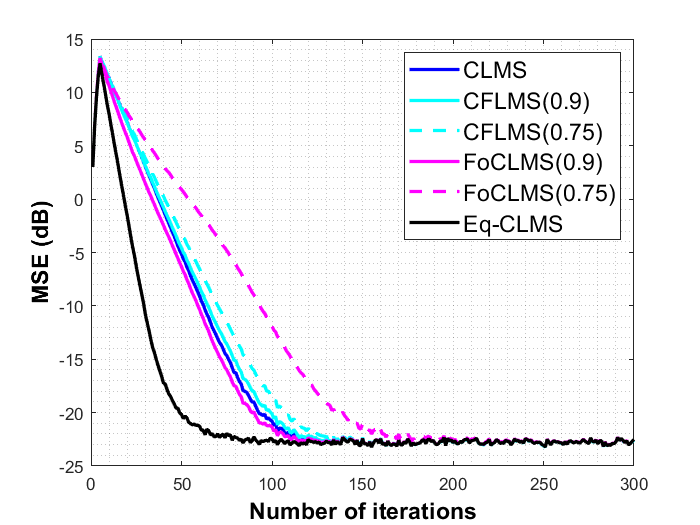}
\caption*{(a)}
\end{minipage}
\begin{minipage}[b]{.7\textwidth}
\includegraphics[width=\textwidth]{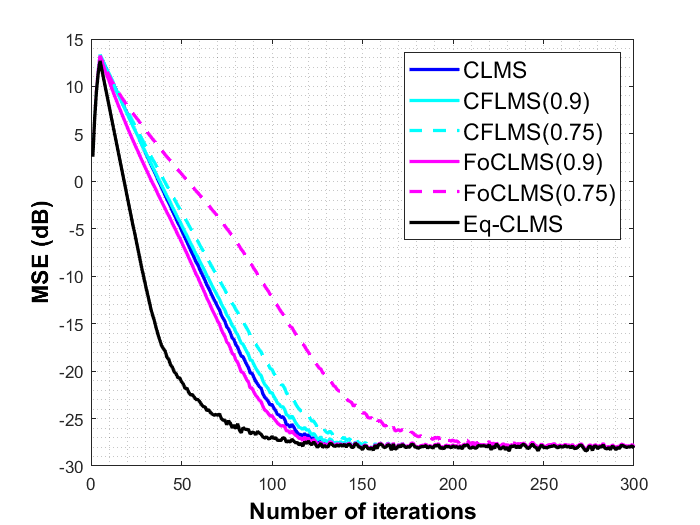}
\caption*{(b)}
\end{minipage}
\caption{MSE curves for the CLMS, CFLMS (f = 0.75 and 0.9), FoCLMS (f = 0.75 and 0.9), and the proposed $Eq$-CLMS algorithm with the SNR of (a) $10$ dB and (b) $20$ dB.}
\label{CE}
\end{figure}

The performance of the proposed E\emph{q}-CLMS algorithm is evaluated for the problem of channel equalization (refer to Figure \ref{Wchannel}).  The transmitted data  $\mathbf{x}(n)$ consisting of $3\times10^{2}$ complex randomly generated samples drawn from the standard Gaussian distribution, is passed through the channel modeled as a finite impulse response (FIR) filter. The size of the channel is taken to be $1 \times 5$ initialized by taking all ones and the weights initialized randomly.  Zero-mean additive white Gaussian noise (AWGN) is added to the channel output.  To cancel the effects of the channel, the noisy signal is sent to the equalizer and the weights of the equalizer are updated through an adaptive algorithm. For the modulation of input signal binary phase-shift keying (BPSK) is used and the error is calculated by taking the difference between the true output (equalized signal) and the desired response (original input signal).  For the performance assessment, mean squared error (MSE) is used given as
\begin{equation}
    \text{MSE} = \frac{1}{n} \sum_{i=1}^{n}(e(n))^{2}
\end{equation}
where $n$ represents the number of data points and $e$ is the error generated between the true and desired signal.

Performance of the proposed E\emph{q}-CLMS approach is compared with three contemporary methods namely: (1) complex LMS (CLMS), (2) fractional CLMS, and (3) fractional order complex LMS (FoCLMS). The experiments are conducted on two different SNR levels of $10$ dB and $20$ dB with step-sizes of $2e^{-3}$, $1e^{-3}$, $2.5e^{-3}$, and $1.5e^{-2}$ for CLMS, CFLMS, FoCLMS, and E\emph{q}-CLMS, respectively. The hyper-parameters of all methods are set to achieve similar steady-state-error.  For fractional variants, two fractional powers are chosen i.e., $0.9$ and $0.75$ and the simulations are run for $300$ iterations, mean results are reported for $1000$ independent runs.  Comprehensive results are tabulated in Table \ref{tab:resultsexp}.  The MSE curves for CLMS, CFLMS, FoCLMS, and the proposed E\emph{q}-CLMS algorithms are shown in Figure \ref{CE}.  The proposed E\emph{q}-CLMS algorithm is shown to have the fastest convergence rate for both values of SNR. For fractional power of $0.75$ and $0.9$, the CFLMS has the slowest convergence rate at both SNR levels. Furthermore, CLMS and FoCLMS show faster convergence when compared to CFLMS, but slower when compared to the proposed technique.  In particular,  the proposed E\emph{q}-CLMS algorithm achieved the fastest convergence at $51^{th}$ and $52^{th}$ iterations for $10$ dB and $20$ dB SNR respectively.  This performance is $2.01$ and $2.11$ times faster compared to the CLMS algorithm for the two SNR levels respectively.  In comparison to the fractional variants of CLMS for two different values of fractional power i.e., $0.9$ and $0.75$, the CFLMS requires $59$ and $70$ more iterations for $10$ dB SNR level.  For $20$ dB SNR it requires $59$ and $75$ more iteration compared to the proposed E\emph{q}-CLMS algorithm.  The FoCLMS algorithm requires $49$ and $98$ more iterations  (corresponding to fractional powers of 0.9 and 0.75 respectively) for $10$ dB SNR and $48$ and $98$ more iterations  (corresponding to fractional powers of 0.9 and 0.75 respectively) for $20$ dB SNR compared to the proposed approach.  Results are summarized in Table \ref{tab:resultsexp}.  These results comprehensively demonstrate the superiority of the proposed algorithm in terms of convergence rate.

\section{Conclusion}\label{Sec:Con}
In this work, a novel adaptive algorithm for complex domain has been presented.  The proposed algorithm, called the \emph{q}-CLMS algorithm, is based on \emph{q}-calculus and makes use of the Wirtinger method.  Through the $q$ parameter, the proposed method offers more control over the steady-state and convergence performances compared to the contemporary approaches.  The transient and steady-state analyses of the proposed algorithm were carried out and theoretical findings are found to be in agreement with the simulation results.  The proposed concept is further automated through an adaptive approach called as enhanced q-CLMS (E\emph{q}-CLMS).  The enhanced approach automatically selects the best value of the 'q' parameter without requiring the hit and trial approach.  Extensive experiments have been conducted on three significant problems namely:  (1) System identification, (2) whitening filter, and (3) channel equalization.  The proposed algorithm has shown superior performance compared to the contemporary approaches.

\section*{Acknowledgements}

This project was funded by the Deanship of Scientific Research (DSR), King Abdulaziz University, under grant No. (DF-209-135-1441). The authors, therefore, acknowledge the technical and financial support of DSR.

\section*{Compliance with ethical standards}
\textbf{Conflict of interest} The authors declared that they have no conflict of interest to this work.

	\bibliographystyle{unsrt}
	\bibliography{References_qCLMS2}

\end{document}